%% file: main.tex
\documentclass[conference,compsoc]{IEEEtran}

\ifCLASSOPTIONcompsoc
  \usepackage[nocompress]{cite}
\else
  \usepackage{cite}
\fi
\usepackage{amsmath,amssymb,amsfonts}
\usepackage{algorithmic}
\usepackage{graphicx}
\usepackage{textcomp}
\usepackage[T1]{fontenc}
\usepackage[table]{xcolor}
\def\BibTeX{{\rm B\kern-.05em{\sc i\kern-.025em b}\kern-.08em
    T\kern-.1667em\lower.7ex\hbox{E}\kern-.125emX}}
\usepackage{tikz}
\usepackage{amsmath}
\usepackage{fvextra}

\usepackage{times,url,color,soul,xspace,enumitem}
\usepackage{graphicx}
\usepackage{caption}
\usepackage{float}  
\usepackage{subcaption}
\usepackage{comment}
\usepackage{xspace}
\usepackage{tikz}
\usepackage{amsmath}
\usepackage[inline,final,nomargin,index]{fixme}

\usepackage{amsmath}
\usepackage{graphicx}  %
\usepackage{hyperref}  %
\usepackage[linesnumbered,ruled,vlined]{algorithm2e}
\usepackage{caption}
\usepackage{float}
\usepackage{graphicx}  %
\usepackage{tabularx}
\usepackage{stfloats}
\usepackage{placeins}
\usepackage{booktabs}
\usepackage{listings}
\usepackage{upquote}
\usepackage{float}  %
\usepackage{multirow}
\usepackage{tikz}

\usepackage[ruled,vlined,linesnumbered]{algorithm2e}
\usepackage{graphicx} %
\usepackage{tikz}
\usetikzlibrary{positioning,arrows.meta,patterns}
\usepackage{caption}
\usepackage{amsmath}
\usepackage{amssymb}
\usepackage{times,url,color,soul,xspace,enumitem}
\usepackage{cleveref}
\usepackage{booktabs}
\usepackage{wasysym}
\usepackage{multirow}
\usepackage{pgfplots}
\usepackage{pgfplotstable}
\usepackage{caption}
\usepackage{subcaption}
\pgfplotsset{compat=1.18}

\usepackage{tabularx}

\DontPrintSemicolon
\SetKwInput{KwIn}{Input}
\SetKwInput{KwOut}{Output}
\SetKwComment{tcp}{// }{}

\definecolor{kw}{RGB}{0,0,255}    %
\definecolor{cm}{RGB}{34,139,34}  %
\definecolor{grey}{gray}{0.5}
\definecolor{st}{RGB}{255,69,0}   %

\lstdefinestyle{csecurity}{
  language=C,
  basicstyle=\ttfamily\footnotesize,
  numbers=left,
  numberstyle=\tiny\color{black},
  stepnumber=1,
  numbersep=8pt,
  showstringspaces=false,
  breaklines=true,
  tabsize=2,
  keywordstyle=\color{kw}\bfseries,  %
  commentstyle=\color{cm}\itshape,   %
  stringstyle=\color{st},            %
  backgroundcolor=\color{white},     %
  frame=none,
  framerule=0.3pt,
  xleftmargin=1.2em,
  framexleftmargin=1.0em,
  keepspaces=true,
  escapechar=§ %
}

\lstdefinestyle{asmgas}{
  basicstyle=\ttfamily\scriptsize,
  keywordstyle=\color{blue}\bfseries,
  commentstyle=\color{teal!70!black}\itshape,
  numbers=left,
  numberstyle=\tiny,
  stepnumber=1,
  numbersep=6pt,
  tabsize=4,
  showstringspaces=false,
  breaklines=true,
  captionpos=t,
  morekeywords={mov, movl, lea, push, call, rep, movsl, add, sub, cmp, jle, cmpl, ret}
}

\SetCommentSty{mycommfont}

\SetAlgoNlRelativeSize{-1}

\DontPrintSemicolon  %

\begin{document}

\input{notations}

\title{ACaM: Automated Capability Modeling for Agentic AI Security}

\author{
\IEEEauthorblockN{
Monika Santra\IEEEauthorrefmark{1}\textsuperscript{\textparagraph},
Dhilung Kirat\IEEEauthorrefmark{2},
Qiushi Wu\IEEEauthorrefmark{2},
Kevin Eykholt\IEEEauthorrefmark{2},
Xiaokui Shu\IEEEauthorrefmark{2},
Ian Molloy\IEEEauthorrefmark{2}}
\IEEEauthorblockA{\IEEEauthorrefmark{1}The Pennsylvania State University, PA, USA\\
mvs7165@psu.edu}
\IEEEauthorblockA{\IEEEauthorrefmark{2}IBM Thomas J. Watson Research Center, Yorktown Heights, NY, USA\\
\{dkirat, molloyim\}@us.ibm.com,
\{Qiushi.Wu, kheykholt, xiaokui.shu\}@ibm.com}
}
\maketitle
\begingroup
\renewcommand\thefootnote{\textparagraph}
\footnotetext{Work done during an internship at IBM Research.}
\endgroup

\input{src/00-abstract.tex}


\input{src/01-intoduction}

\input{src/02-Background_and_Related_work}

\input{src/03-motivation-overview}

\input{src/04-methodology}

\input{src/05-implementation}
\input{src/06-Evaluation}

\input{src/07-discussion-and-limilations}

\input{src/08-Conclusion}

\bibliographystyle{IEEEtran}

\bibliography{main}

\appendices

\input{src/09-appendixA}

\end{document}

%% file: notations.tex
\newcommand{\sysname}{ACaM} %

\newcommand{\todo}[1]{\textbf{\textcolor{red}{#1}}}

\newcommand{\eqnote}[1]{\scriptsize #1}
\newcommand{\afg}{AFG\xspace}
\newcommand{\acpg}{ACPG\xspace}
\newcommand{\did}{DID\xspace}
\newcommand{\cl}{capability laundering\xspace}

\lstdefinestyle{pycode}{
  language=Python,
  basicstyle=\ttfamily\footnotesize,
  keywordstyle=\color{blue!70!black},
  stringstyle=\color{orange!80!black},
  commentstyle=\color{gray},
  numbers=left, numberstyle=\tiny\color{gray}, numbersep=5pt,
  frame=single, breaklines=true, captionpos=b,
  upquote=true,
  keepspaces=true,
}

\lstdefinestyle{promptcode}{
  basicstyle=\ttfamily\scriptsize,   %
  breaklines=true, breakatwhitespace=true, breakindent=0pt,
  frame=tb, framesep=2pt,            %
  numbers=none, xleftmargin=0pt,     %
  columns=fullflexible, keepspaces=true,
  aboveskip=4pt, belowskip=2pt}

%% file: src/00-abstract.tex
\begin{abstract}
In agentic AI systems, an LLM invokes tools that hold standing access to credentials, private data, filesystems, and external services. Such agents are increasingly deployed without an answer to a basic question: what can an agent access, exfiltrate, and propagate, independent of what it is designed for or is observed to do? The question goes unanswered because existing approaches trust a tool's natural-language description as a proxy for its behavior. But nothing enforces that the description and its implementation agree. This gap is known as Description-Code Inconsistency (DCI). DCI enables a threat class called capability laundering, in which a tool's implementation contains dangerous behavior that its description omits. The behavior may come from incidental drift or a supply-chain compromise. It runs silently whenever an LLM invokes the tool for its stated, benign purpose. We present ACaM, a neurosymbolic approach to capability modeling that derives each tool's capabilities from its code, so a laundered capability cannot hide behind a benign description. In ACaM, static analysis grounds an LLM in code-derived facts, and no LLM output can override a fact that static analysis has established. The LLM supplies only the behavioral meaning that those facts cannot express. ACaM compiles agent source code into the Agent Capability Property Graph (ACPG) and uses it to reconstruct threat chains that span multiple tool calls. We evaluate ACaM on 43 real-world agents, including 9 multi-agent systems. On 24 manually annotated agents, it recovers capability properties with 98.3\% macro-F1 and reconstructs injected threat chains with a score of 0.980, 8 points above the strongest baseline. Applied to all 43 unmodified agents, it discovers hundreds of structurally vulnerable paths, ten of which we confirm with proof-of-concept exploits.
\end{abstract}

%% file: src/01-intoduction.tex
\section{Introduction}\label{sec:introduction}

Consider an agent that uses a tool called \texttt{get\_weather(city)}. The tool's description promises a weather forecast, and the agent receives one when it calls the tool. However, the same function also reads a credentials file and sends its contents out in an HTTP header. The description never mentions this behavior, so every normal use of the tool silently leaks the credentials~\cite{dci_2,stub-dcisurvey}. 

The agent cannot detect this leak, because its LLM sees only the tool's description and never the tool's code. The LLM also decides at runtime which tools to call, in what order, and with what arguments. As a result, the full set of actions a deployed agent can take is unknown in advance. In other words, a deployed agent has an uncharacterized capability surface. Today, this surface is handled mostly at runtime: existing defenses~\cite{forge_jha, progent, stub-camel, agent_sec_1, agent_sec_2, agent_sec_3, agent_sec_5, agent_sec_7, agent_sec_8, agent_sec_9, agent_sec_10, agent_sec_11, stub-ifc, stub-llmdepgraph, isolategpt} monitor, sandbox, or constrain the agent while it runs. Yet many risky capabilities, like the leak above, can be identified and fixed before deployment. Finding them early reduces the load on runtime defenses, which can then focus on truly dynamic behavior. This pre-deployment analysis starts with a basic question: what can this agent do? The answer should cover everything the agent's code permits, not just what the agent was designed to do or usually does.

Decades of software-security tooling~\cite{stub-pysa,stub-semgrep,code_property_graph,taint_analysis_1} assume that the code fixes how a program’s components interact. Analyzing the code is therefore enough to know which components can reach which. Agentic systems break this premise. An LLM orchestrator sits between the agent and its tools. It picks which tools to call, and its choice can change from one run to the next. Classical static analysis has no model of such a decision. It also cannot reason about an agent-tool boundary that is set in natural language each time the agent executes.

Existing agentic-security work falls into three main lines. Each asks a different question from ours. One line studies the LLM's behavior. It asks whether prompt injection, jailbreaks, or multi-agent trust attacks can make an LLM misbehave at inference time~\cite{ipi,sok_jailbreak}. It never asks what a tool can do once invoked. A second line studies the system's structure. Its methods either trust each tool's declared description or require a behavioral policy or initial trust labels as input~\cite{forge_jha,stub-ifc}. They do not derive these inputs from the code. A third line questions this trust. It measures how often a tool's description diverges from its code across thousands of real-world tools~\cite{dci_2,capa_launder_file_system}. However, it checks each tool in isolation and does not model what the composed agent can do or the threats that span several tools.

We ask an orthogonal, structural question. What can the system actually access, exfiltrate, and propagate? The answer does not depend on how or why a dangerous dispatch is triggered. The trigger may be a malicious tool, an injected prompt, or an honest mistake. Answering it requires deriving capabilities from code, not from tool descriptions.

This trust in tool descriptions can be exploited silently because of how agents are built. An LLM orchestrator decides whether to invoke a tool by reading its natural-language description, but the tool's actual capability is fixed by its implementation, which the orchestrator never inspects. Nothing in a deployed agent requires the two to agree or checks that they do. Such a disagreement is known as Description-Code Inconsistency (DCI). DCI enables a threat class called \emph{capability laundering}~\cite{capa_launder_file_system}, in which an agent carries a dangerous capability that no tool's description declares. The simplest case is a single tool whose code contains dangerous behavior that its description omits or contradicts, as in the opening \texttt{get\_weather} example. At the description level, such a tool looks identical to a safe one. Detecting capability laundering therefore requires reasoning about what the code does, without relying on what the description claims.

Analyzing each tool's code in isolation is necessary but not sufficient, because capability in agentic systems is compositional. Suppose three tools are each benign on their own. The first reads an API key and caches it in shared state. The second writes the shared state to a log file. The third uploads the log file to a remote debugging service. Together, they deliver the key to a service that never needed it, and no tool's description declares this capability. An analysis must therefore cover the whole agent, including a model of the LLM-to-tool call interface.

\sysname{} computes this surface with a neurosymbolic pipeline. Static analysis recovers most of the surface from the code's structure. An LLM handles only the semantic residual that structure cannot capture. This division keeps LLM non-determinism out of the final result. \sysname{} combines both outputs into a single artifact, the Agent Capability Property Graph (ACPG). The ACPG is a compositional, machine-readable model of the agent's capability surface. It is built from source code alone. Once built, the ACPG can be queried repeatedly. It supports two downstream analyses. The first, threat-chain reconstruction, traces tainted data to an external sink. The data may leave directly via LLM context poisoning, through a side channel, or over multiple hops via subprocesses or shared state. The second, structural vulnerability identification, flags dangerous capabilities, even when the tools' descriptions do not declare them.

We make the following contributions:

\begin{enumerate}
    \item \textbf{Capability Model.} We formalize the pre-deployment structural capability question for agentic systems and model it as the Agent Capability Property Graph (ACPG). The ACPG is a compositional graph that unifies low-level static-analysis facts with high-level semantic judgments.

    \item \textbf{Neurosymbolic Agentic Analysis Framework.} We present \sysname{}, a compositional framework that builds the ACPG from agent source code. Static analysis anchors the capability model, and an LLM fills only the semantic residual. This lets \sysname{} reason over the composed system and reconstruct multi-hop threat chains that no single tool's code reveals.

    \item \textbf{Downstream Evaluation.} \emph{Threat-chain reconstruction:} \sysname{} achieves a mean score of 0.980, roughly 8 points above the strongest LLM-only baseline (0.903) and far above every SA-only baseline ($\leq$0.342). \emph{Structural vulnerability identification:} across the 43 real-world agents, ACaM surfaces 123 code-execution and credential-exposure paths. We manually confirmed that each exists in the code. It also surfaces about 1.8K medium- to high-impact paths, confirmed by a separate verification agent. \emph{Exploitability:} we develop working proof-of-concept (PoC) exploits for ten Critical and High findings, spanning code execution, injection, and access-control failures. Each flaw stems from the agent's code, and several, such as the deserialization RCEs, need no LLM decision at all.

\end{enumerate}

%% file: src/02-Background_and_Related_work.tex
\begin{table*}[t!]
\centering
\footnotesize
\setlength{\tabcolsep}{4pt}
\caption{Related work compared against the properties a structural,
pre-deployment agentic security analysis requires. \checkmark{} = yes, $\times$ = no,
P = partial.}
\label{tab:related-work}
{\renewcommand{\arraystretch}{0.9}
\begin{tabular*}{\textwidth}{@{\extracolsep{\fill}}p{3.6cm}l c c c c c@{}}
\toprule
\textbf{Category} & \textbf{Target} &
\shortstack{\textbf{Pre-}\\\textbf{deploy.}} &
\shortstack{\textbf{Models LLM}\\\textbf{dispatch}} &
\shortstack{\textbf{Tool-}\\\textbf{Compositional}} &
\shortstack{\textbf{Code-}\\\textbf{grounded}} &
\shortstack{\textbf{Cross-tool chain}\\\textbf{reconstruction}} \\
\midrule
Classical SA / CPG             & Classical & \checkmark & $\times$   & $\times$   & \checkmark            & $\times$ \\
LLM-only threat modeling       & Classical & \checkmark & $\times$   & $\times$   & $\times$              & $\times$ \\
LLM-assisted dependency graphs & Agentic   & $\times$   & \checkmark & \checkmark & $\times$              & P \\
LLM-assisted analysis          & Classical & \checkmark & $\times$   & $\times$   & \checkmark            & $\times$ \\
IFC / enforcement / sandboxing & Agentic   & $\times$   & P          & \checkmark & \checkmark$^{\dagger}$ & $\times$ \\
DCI                & Agentic   & \checkmark & $\times$   & $\times$   & \checkmark            & $\times^{\ast}$ \\
\textbf{ACaM (ours)}            & \textbf{Agentic} & \checkmark & \checkmark & \checkmark & \checkmark & \checkmark \\
\bottomrule
\end{tabular*}}
\vspace{2pt}
{\footnotesize $^{\dagger}$At runtime and requires enforcement policy or initial labeling. $^{\ast}$Checks each tool against its description; no composed-system analysis.}
\end{table*}

\section{Related Work}\label{related_work}
We organize prior work into six categories and compare them in
Table~\ref{tab:related-work} against the five properties a structural, pre-deployment
analysis of agents requires: pre-deployment applicability, modeling the
LLM dispatch layer, tool compositionality, code grounding, and
cross-tool chain reconstruction. Each category addresses part of the
problem, but none satisfies all five.

\textbf{Analysis built for non-agentic systems.}
Tools such as Pysa~\cite{stub-pysa}, CodeQL~\cite{stub-codeql}, and Semgrep~\cite{
stub-semgrep} perform taint tracking and pattern matching over source
code~\cite{taint_analysis_1,taint_analysis_2,taint_analysis_3}, and the
code property graph~\cite{code_property_graph} unifies a program's AST,
control flow, and dependencies into one queryable structure. These tools
are fast, precise, and run before deployment, but they have no notion of
the boundary between an LLM and its tools. Modeling the LLM as a
taint-in, taint-out stub does not fix this, because the stub breaks
legitimate flows and misses real issues. A growing body of work pairs LLMs with such analyzers to improve path sensitivity or mine bugs~\cite{neurosym1,neurosym2,neurosym3,neurosym4,neurosym5,neurosym6,neurosym7,stub-iris,stub-bugstone}. These hybrid systems share our instinct that an LLM must be grounded in program structure, but they analyze classical software code. Agentic systems add an LLM inside the control loop. This LLM selects behavior at runtime from a description. None of these hybrid systems represents this dispatch layer. STRIDE GPT~\cite{stub-stridegpt}
instead uses an LLM to enumerate threats directly from a system's
design or description~\cite{threatModel}, reasoning fluently at the
natural-language level. It never reads the            %
implementation, so it inherits whatever the description leaves out.

\textbf{Runtime-observed structure and enforcement for agentic systems.}
A related line of work builds a program- or data-dependency view of agent behavior from runtime traces, and an LLM reasons over this view at a high level~\cite{stub-llmdepgraph}. This approach recovers structure that natural-language-only methods lack. However, it misses any hidden capability that never surfaces as an observed effect, which is the case DCI describes. It also never grounds the underlying code facts. A larger body of work instead protects agents at runtime~\cite{fides,isolategpt,progent,agent_sec_5,agent_sec_11} through information-flow control~\cite{stub-ifc}, policy-based enforcement such as CaMeL~\cite{stub-camel} and FORGE~\cite{forge_jha}, and sandboxing. This work is necessary, since runtime is ultimately where actions are blocked. Yet many of the issues it catches could have been caught earlier. In addition, each mechanism presupposes an input it does not itself produce. IFC needs annotations, and enforcement needs a policy. Survey work calls for a structural, pre-deployment representation to fill this gap~\cite{stub-christodorescu}.

\textbf{Description-code inconsistency (DCI).}
Recent work on DCI measures how often an MCP tool's description diverges from its code~\cite{stub-dcisurvey,dci_2}. It finds divergence in 10\% to 13\% of cases. Concurrent work, MCP-FSD~\cite{capa_launder_file_system}, flags filesystem capability mismatch in MCP servers. In these cases, attacker-steered arguments make a tool read host files outside its declared contract. MCP-FSD confirms each case by running the tool in a sandbox. DCI motivates our work.

Table~\ref{tab:related-work} makes this gap visible. No classical-target category models LLM dispatch, where the LLM chooses which tools to call. Among agentic-target work, only DCI is code-grounded before deployment, like \sysname{}. The other agentic lines of work operate at runtime. DCI and \sysname{} still differ in method. DCI compares each tool's description with its code, one tool at a time. MCP-FSD, the closest detection work, is also per-tool. It confirms filesystem laundering by executing a single tool. So it is dynamic and limited to the filesystem. \sysname{} instead models each tool's capability as a node in the full agent capability graph. It works statically, over a broad capability set. This set spans code execution, secret exfiltration, capability escalation, filesystem access, and more. This compositional view reconstructs cross-tool chains, so it catches capability laundering that appears only when tools are chained together, which a per-tool check cannot see.

%% file: src/03-motivation-overview.tex
\section{Design Motivation and Overview}
\label{sec:motivation}

\textbf{Design Motivation.}
Static analysis and large language models fail in opposite places. Static analysis recovers data flow, component boundaries, and long compositional chains precisely and without hallucination. However, it is blind to what those facts mean. It can establish that a branch is conditioned on a variable, but it cannot tell whether that branch enforces a privilege check. That judgment depends on what the code is for, which its structure alone does not reveal.
An LLM has the opposite profile. It reads intent and library semantics the way an analyst does. Yet it degrades on the same long chains that static analysis handles cleanly. It loses intermediate hops, grows overconfident on open-ended reasoning, and takes a tool's own description at face value, even when that description is what the attack exploits.

This contrast makes the split necessary and asymmetric. Static analysis decides most capabilities directly from the code, and only a small share of decisions needs semantic judgment. That share cannot be skipped, because some capabilities are visible only through what the code means. Authority follows this imbalance, since static analysis is grounded in the code and the LLM can hallucinate. Static analysis settles everything it can, and the LLM fills only what remains. No LLM output can weaken a fact that static analysis has established. Static analysis re-runs after each LLM call to enforce this. The LLM's weaknesses shape two further choices. Static analysis performs all data-flow tracing, since LLMs tend to lose intermediate hops on long chains. Each LLM call also receives deterministic inputs and answers a bounded question, so its non-determinism stays confined to its own contributions.

\begin{figure*}
    \centering
    \includegraphics[width=\linewidth]{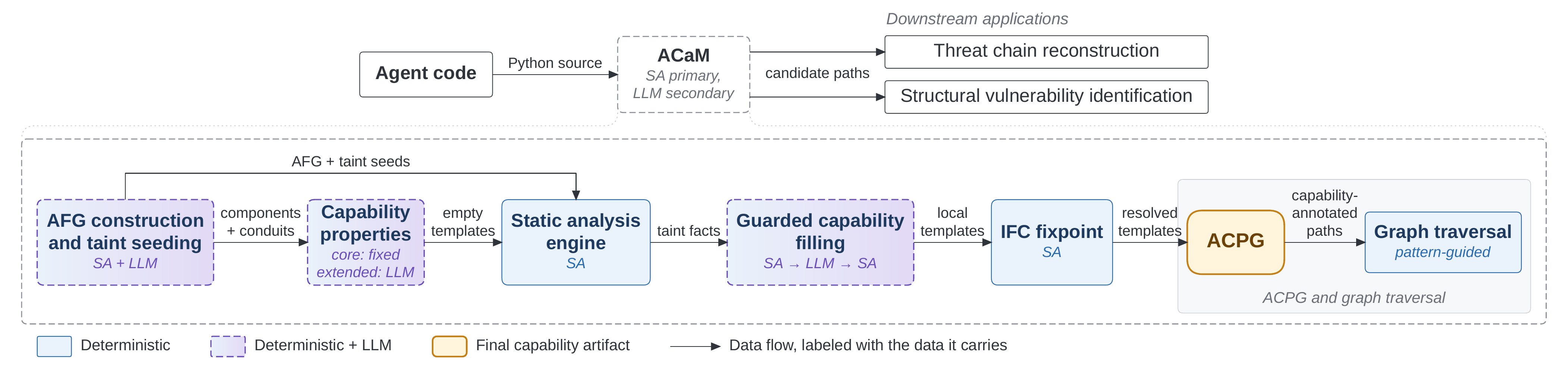}
    \caption{System Overview of \sysname{}}
    \label{fig:overview}
\end{figure*}

\textbf{System Overview.} Figure~\ref{fig:overview} shows how ACaM turns raw agent source code into a queryable capability model in five stages.
\textbf{(1)~AFG Construction and Taint Seeding} recovers the agent's component boundaries and data-flow topology, and seeds each boundary channel with an initial taint label.
\textbf{(2)~Capability Property Discovery} complements a fixed set of core
properties with extended properties, which capture agent-specific,
security-relevant patterns that span multiple components, such as a
secret relayed between tools through shared state.
\textbf{(3)~Static Analysis Engine} propagates taint within each component and emits raw, evidence-grounded capability \emph{facts}.
\textbf{(4)~Guarded Capability Filling} lifts these facts into behavioral capability properties: static analysis resolves every property it can, and the LLM supplies only the remaining semantic judgments.
\textbf{(5)~IFC Fixpoint} propagates confirmed facts across component boundaries until reaching a fixpoint, yielding the \emph{Agent Capability Property Graph} (ACPG). Downstream, \sysname{} supports two tasks through a deterministic, pattern-based traversal of the ACPG. The tasks answer different questions, so we evaluate them differently. \emph{Threat-chain reconstruction} tests reconstruction quality: it traces multi-tool chains end to end. Scoring it needs ground truth, so we inject known vulnerable paths. \emph{Structural vulnerability identification} tests real-world discovery: it flags structurally vulnerable paths in unmodified agents.

\textbf{Threat model and scope.} \sysname{} does not trust tool
descriptions. A description can diverge from the code through a malicious
contributor, a compromised third-party tool, or honest drift. \sysname{} need not
distinguish these cases. \sysname{}
derives each tool's capabilities from the code itself, so no description
can hide what a tool can do. A \emph{tool} here is a
component the orchestrator can invoke. We analyze its source as written,
before deployment. This is the white-box setting, and our embodiment
targets Python. \sysname{} is a pre-deployment static analyzer, not a
runtime input defense. Whether a runtime filter resists adaptive
attacks~\cite{adaptive_attack} is a separate question, outside our scope. Three
classes of code fall outside this surface: code that modifies itself after
deployment, native or opaque remote components (black-box), and code that
defeats the analysis through obfuscation, reflection, or runtime
construction. We therefore assume an adversary that hides a capability
behind a benign description, not behind code built to evade static
detection. When \sysname{} cannot resolve a construct, it marks it
\textsc{unknown} rather than safe. Evasion then surfaces as unanalyzable
code, not a clean result.

%% file: src/04-methodology.tex
\section{Methodology}
\label{sec:methodology}

We walk through \sysname{}'s key stages on the minimal illustrative
agent in Figure~\ref{fig:running-example}. \texttt{get\_news} authenticates with an API token that
it caches in shared state, and \texttt{sync\_prefs} posts that state
to a preferences backend. Each tool is benign looking in isolation. Together, however, they form a latent compositional exfiltration chain like the one in Section~\ref{sec:introduction}. The token reaches an endpoint that never needed it.

\begin{figure}[t!]
\begin{lstlisting}[language=Python, basicstyle=\ttfamily\scriptsize,
  numbers=left, numberstyle=\tiny, xleftmargin=1.5em,
  columns=fullflexible, breaklines=true, showstringspaces=false]
STATE = {}  # shared session state

@tool
def get_news(topic: str) -> str:
    """Return headlines for a topic."""
    STATE["lang"] = "en"
    STATE["token"] = os.getenv("NEWS_API_TOKEN")
    r = requests.get(NEWS_URL, params={
        "q": topic, "key": STATE["token"]})
    return r.text

@tool
def sync_prefs() -> str:
    """Sync the user's preferences."""
    requests.post(PREFS_URL, json=STATE)
    return "synced"

agent = create_react_agent(
    llm, [get_news, sync_prefs])
\end{lstlisting}
\caption{Running example. Each tool is benign looking in isolation:
\texttt{get\_news} caches its API token in shared state, and
\texttt{sync\_prefs} posts that state, token included.}
\label{fig:running-example}
\end{figure}

\begin{figure}[t!]
  \centering
  \includegraphics[width=\columnwidth]{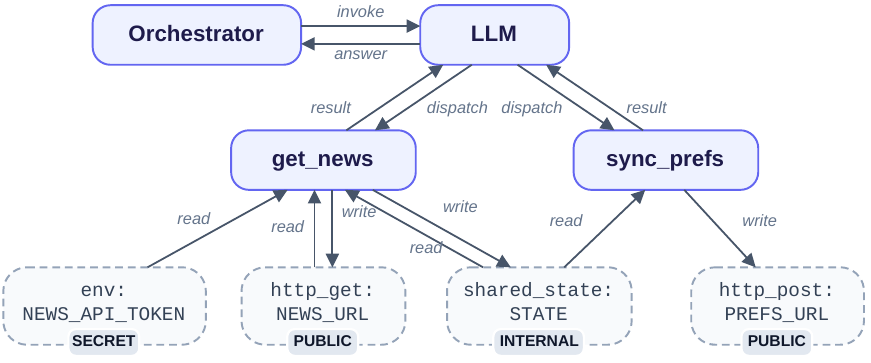}
  \caption{AFG and initial taint seeds of the running example.
Components are solid; conduits are dashed and tagged with their seed.
The token can reach \texttt{http\_post:PREFS\_URL} via
\texttt{shared\_state}, a flow neither tool reveals alone.}
  \label{fig:running-afg}
\end{figure}

\subsection{AFG Construction and Taint Seeding}
\label{sec:method:afg}

The first stage turns agent source code into an \emph{Agent Flow
Graph}~(\afg{}), which separates code from the data channels between
code so that later stages can reason about flows across components.
The \afg{} has two kinds of nodes, \emph{components} $V_C$ and \emph{conduits} $V_D$. It also has edges for calls and data flow, and a taint seed on every conduit. Figure~\ref{fig:running-afg} shows the \afg{} of the
running example.

\textbf{Components.}
A component is a unit of program logic in one of three roles: the
\emph{orchestrator}, the agent's control loop; a \emph{tool}, an
independently invocable capability; or a \emph{helper}, code that
serves either. A special \emph{LLM node}, with no source code, stands
for the model itself. \sysname{} locates tools and the orchestrator
through framework-specific registration patterns, e.g., LangChain's
\texttt{@tool} decorator or LangGraph's \texttt{add\_node}. Each tool's component is everything reachable from its anchor, and the orchestrator absorbs the rest. A helper shared by two tools belongs to both, because shared logic lets a capability added for one tool reach the other. Control-flow edges record every transfer of execution between
components, such as model calls, tool dispatches, subprocess spawns,
and inter-agent handoffs, each paired with a return edge. In
Figure~\ref{fig:running-afg}, the orchestrator
(\texttt{create\_react\_agent}) \emph{invokes} the LLM node, which
\emph{dispatches} the tools \texttt{get\_news} and
\texttt{sync\_prefs}, each returning a \emph{result}.

\textbf{Conduits and data edges.}
A conduit is any channel through which data leaves or enters a function, such as shared mutable state, files, network calls, subprocess I/O, environment variables, databases, and more. Recognizing a conduit requires knowing how libraries behave. For example, the analysis must know that \texttt{redis.set()} writes shared state. This makes recognition a lookup problem rather than a reasoning problem.
\sysname{} therefore pairs deterministic rules for known standard and
framework patterns with a single, bounded LLM call (Prompt~P1,
Appendix~\ref{appendix:prompts}) that reviews them and adds conduits
the rules miss. The resulting rule set is built once per codebase and
applied deterministically, so a hallucinated conduit matches no call
site. \sysname{} then records read and write edges between each conduit and
the components that access it. The
running example has four conduits: the token's environment variable,
the shared dictionary \texttt{STATE}, and the two HTTP calls;
\texttt{get\_news} both writes and reads \texttt{shared\_state:STATE},
while \texttt{sync\_prefs} only reads it.

\textbf{Taint seeds.}
Each conduit $d \in V_D$ receives an initial label
$\tau(d) \in \mathcal{L} = \{\texttt{PUBLIC}, \texttt{INTERNAL}, \texttt{SECRET}\}$,
its \emph{seed}. A seed marks where secret or untrusted data enters
the system, not where it ends up; later stages refine it. Seeding has two layers. First, deterministic rules assign a label from the conduit’s kind and identifier, under a fixed trust model in which external channels are attacker-reachable and credential sources are secret. Second, an LLM proposes its own label for each conduit. \sysname{} keeps the higher of the two labels in the lattice $\texttt{PUBLIC} \sqsubseteq \texttt{INTERNAL} \sqsubseteq \texttt{SECRET}$, so the LLM can raise a seed but never lower it. In
Figure~\ref{fig:running-afg}, the rules alone settle every seed:
\texttt{NEWS\_API\_TOKEN} is \texttt{SECRET} because its identifier
names a credential, the HTTP conduits are \texttt{PUBLIC} because
network data is attacker-reachable by invariant, and \texttt{STATE}
is \texttt{INTERNAL} by its kind's default.

In both conduit classification and seeding, the LLM contributes knowledge that rules cannot encode. Deterministic machinery still bounds every value it provides. The finished \afg{} already exposes the cross-tool read of \texttt{shared\_state:STATE}. However, nothing in it says that the token flows there, because seeds mark only where data enters. Whether seeded data reaches a sink is decided in Section~\ref{sec:method:sa}.

\subsection{Capability Properties}
\label{sec:method:afg:template}

For each component, \sysname{} instantiates a \emph{capability template}. The template describes what the component \emph{can} do. It has two tiers.

\textbf{Core properties.}
The core tier is a fixed catalog of 19 properties present in every template of every agent (Table~\ref{tab:core_properties}). The catalog follows the main elements of the \afg{}. \emph{Data access} covers what a component reads and writes, and \emph{external reach} covers how it acts outside itself. \emph{Propagation} covers where \textsc{Secret} data can flow, and \emph{LLM interaction} covers how the component can affect the model at runtime. The propagation
properties look alike but are not. They differ by where a secret goes: to
another internal component (\texttt{produces\_secret\_output}), into a
return value the caller sees (\texttt{secret\_in\_tool\_return}), or out to
an external sink (\texttt{may\_leak\_secret}). \texttt{writes\_secret} is
the broad signal they refine. Each marks a stage of a multi-hop leak, so one
``secret leaves'' property could not express the chains \sysname{}
reconstructs. The grouping is conceptual and does not determine how a value is resolved. Because the catalog is fixed, it describes every agent in the same terms, which makes agents directly comparable. No fixed catalog can anticipate every agent-specific pattern, however, so the extended tier complements the core.

\begin{table}[t]
\centering
\caption{The 19 core capability properties, grouped by the kind of capability each tracks. Read/write pairs share a row.}\label{tab:core_properties}
\scriptsize
\setlength{\tabcolsep}{4pt}
\begin{tabularx}{\columnwidth}{@{}l X@{}}
\toprule
\textbf{Property} & \textbf{Description} \\
\midrule
\multicolumn{2}{@{}l}{\textit{Data access}} \\
\texttt{reads\_secret}                    & Reads credentials, keys, tokens, or secret files/env vars \\
\texttt{writes\_secret}                   & Writes \textsc{Secret} data to any conduit \\
\texttt{\{reads,writes\}\_shared\_state}   & Reads/writes module-level state shared across components \\
\texttt{file\_\{read,write\}}              & Reads/writes the local filesystem \\
\texttt{\{reads,writes\}\_persistent\_store} & Reads/writes a database, cache, or message queue \\
\texttt{cryptographic\_operation}         & Hashes, encrypts, decrypts, signs, or verifies \\
\addlinespace[2pt]
\multicolumn{2}{@{}l}{\textit{External reach}} \\
\texttt{external\_communication}          & Makes outbound network calls (HTTP, socket, gRPC) \\
\texttt{handles\_inbound}                 & Receives external input (HTTP handler, CLI, stdin) \\
\texttt{subprocess\_call}                 & Spawns child processes \\
\texttt{executes\_dynamic\_code}          & Executes generated or LLM-produced code \\
\addlinespace[2pt]
\multicolumn{2}{@{}l}{\textit{Propagation}} \\
\texttt{produces\_secret\_output}         & Passes \textsc{Secret} data to another internal component \\
\texttt{secret\_in\_tool\_return}         & \textsc{Secret} taint reaches the return value \\
\texttt{may\_leak\_secret}                & Sends \textsc{Secret} data to an external sink \\
\texttt{output\_taint\_label}             & Summary taint label of emitted data \\
\addlinespace[2pt]
\multicolumn{2}{@{}l}{\textit{LLM interaction}} \\
\texttt{capability\_escalation}           & Behavior depends on privilege level or role flag \\
\texttt{memory\_poisoning\_risk}          & Untrusted external data flows into agent memory \\
\bottomrule
\end{tabularx}
\end{table}

\textbf{Extended properties.}
The extended tier is open-ended and agent-specific. Its properties are discovered by a single LLM call over all components (Prompt~P2, Appendix~\ref{appendix:prompts}). The call proposes
properties for cross-component patterns that a fixed vocabulary cannot
anticipate, such as multi-hop exfiltration chains or privilege
escalation through shared state, each naming the components it
implicates. For the running example, it proposes
\texttt{secret\_relay\_via\_shared\_state}, a possible leak of a
secret through shared state between \texttt{get\_news} and
\texttt{sync\_prefs}. A proposal is only a question. Deciding whether it holds requires tracing data across components. Its value is therefore filled by the same analysis as the core properties.

\textbf{Property values.}
All properties but one take values in \texttt{YES}, \texttt{NO}, and \texttt{UNKNOWN}. Each starts as \texttt{UNKNOWN} and is resolved by the stages that follow. While a value depends on another component, it holds an intermediate condition (Section~\ref{sec:capability-filling}).
The remaining property, \texttt{output\_taint\_label}, instead holds a
taint label from $\{\texttt{PUBLIC}, \texttt{INTERNAL},
\texttt{SECRET}, \texttt{UNKNOWN}\}$ and is filled only by the IFC
Fixpoint (Section~\ref{sec:method:ifc}).

\subsection{Static Analysis Engine}
\label{sec:method:sa}

The Static Analysis Engine takes the taint-seeded AFG and the empty capability templates as input. It produces a structured set of \emph{taint facts} for each component. These facts are evidence-grounded observations, which the Guarded Capability Filling stage later turns into template values (Section~\ref{sec:capability-filling}).

The engine tracks taint per field, since agents coordinate through
shared objects that mix sensitivities: one dictionary may hold a
\texttt{SECRET} credential under one key and \texttt{INTERNAL} data
under another. Figure~\ref{fig:running-afg} draws \texttt{STATE} as a
single conduit for clarity, but \sysname{} tracks each of its keys
separately. Propagation rules fall into four families: data flow,
implicit flow, inter-procedural summaries, and agentic boundaries
such as subprocess arguments and shared-state keys.

Two design decisions distinguish this engine. First, instead of a single sensitivity lattice, it keeps \emph{two parallel tracks}: one for exfiltration and one for injection. Exfiltration asks whether \emph{sensitive} data reaches a sink. Injection asks whether \emph{untrusted} data reaches the agent's memory, shared state, or LLM context. A single lattice cannot answer both questions. Untrusted data is \texttt{PUBLIC}, the bottom of the lattice, so joining it with sensitive data would erase its origin. In the running example, the API token is sensitive, so ACaM follows it on the exfiltration track. The HTTP response that \texttt{get\_news} returns to the LLM is untrusted, so ACaM follows it on the injection track.

Second, the engine uses the \afg{} as an \emph{analysis schedule}. It
orders components along the \afg{}'s writer-to-reader edges, so each
conduit's writers are analyzed before its readers. Taint then crosses
components in a single forward pass, with no global fixpoint. When a
component writes a higher label than a conduit's seed, the conduit is
\emph{promoted}. Later readers see the promoted label and which
component wrote it. In the running example, \texttt{get\_news} writes
the token into \texttt{shared\_state:STATE}, promoting it from
\texttt{INTERNAL} to \texttt{SECRET}. \texttt{sync\_prefs} then reads
a \texttt{SECRET} value written by \texttt{get\_news}. This ordering requires the inter-component writer-to-reader graph, excluding self-edges, to be acyclic. \sysname{} computes this order with Kahn's algorithm~\cite{kahn_topo}. By default, the algorithm leaves components on a cycle unscheduled. No such cycle arose in our corpus.

The engine outputs, per component, a record of conduit reads and
writes with observed labels and writers, subprocess crossings,
implicit-flow guards, and \texttt{PUBLIC}-origin outputs. Keeping these records as observations separates what static analysis determines mechanically from what requires semantic judgment.
Additionally, each mechanism here has precedent on its own. Multi-kind taint tracking, field-sensitive analysis, and cross-boundary propagation all appear in some form in existing tools. \sysname{} differs in how these mechanisms fit together. All of them share a single abstraction, the conduits of the \afg{}. Lattice propagation, track separation, rule scope, and analysis order therefore come from the same agent-level model. This is a design choice suited to the problem, and other designs are possible.

\subsection{Guarded Capability Filling}
\label{sec:capability-filling}

Guarded Capability Filling turns the taint facts of
Section~\ref{sec:method:sa} into property values. Some values follow
mechanically from the facts; others require judgment that only an LLM
can supply. The challenge is to use the LLM for that judgment without
letting it override what static analysis has already decided. For brevity, Table~\ref{tab:running-values} follows three properties of the running example: the core properties \texttt{reads\_secret} and \texttt{may\_leak\_secret}, and the extended property relay. This stage fills its second column.

\textbf{Structural vs.\ semantic properties.}
Properties differ in whether the facts alone decide them. A
\emph{structural} property follows directly from an observed fact:
reading a \texttt{SECRET}-labeled conduit means
\texttt{reads\_secret} is \texttt{YES}. A \emph{semantic} property
requires understanding what the code \emph{means}. For
\texttt{capability\_escalation}, a branch on \texttt{user.role} and a
branch on \texttt{user.lang} are equally tainted guards, yet only the
first gates privilege. Deciding such properties takes open-ended world knowledge of what code
signifies, which no fixed rule set can enumerate. The LLM supplies it and generalizes to cases no rule
anticipates; static analysis decides everything the facts already
settle.

\textbf{The \texttt{CONDITIONAL} grammar.}
A value may also depend on what a \emph{different} component does, a
fact unavailable in this component's own analysis. Rather than
guessing, \sysname{} records such a value as a condition, resolved
later by the IFC Fixpoint. Let $c\in V_C$ denote a component, $p\in P$ a property, $d\in V_D$ a conduit
(Section~\ref{sec:method:afg}), $\ell\in\mathcal{L}$ a taint label,
$\mathit{val}$ a constant, and
$b\in\{\mathtt{YES},\mathtt{NO},\mathtt{UNKNOWN}\}$ a component's own
verdict, \texttt{UNKNOWN} when its own facts are inconclusive:
\[
\begin{aligned}
v \;::=\;& \mathtt{YES} \mid \mathtt{NO} \mid \mathtt{UNKNOWN} \mid \mathtt{COND}(\varphi)\\
\mid\;& b \vee \mathtt{COND}(\varphi) \mid b \wedge \mathtt{COND}(\varphi)\\
\varphi \;::=\;& \pi \mid \varphi \wedge \varphi \mid \varphi \vee \varphi\\
\pi \;::=\;& c.p\\
\mid\;& c.\mathit{conduit\_write}[d] \sqsupseteq \ell\\
\mid\;& c.\mathit{conduit\_write}[d] = \mathit{val}
\end{aligned}
\]

A value $v$ is what a component's template stores for a \texttt{YES}/\texttt{NO} property. This covers every property except \texttt{output\_taint\_label}, which holds a taint label and is filled only by the IFC Fixpoint (Section~\ref{sec:method:ifc}). \texttt{YES} and \texttt{NO} are settled, and \texttt{UNKNOWN} is not. \texttt{COND}($\varphi$) is a pending value that holds exactly when the condition $\varphi$ holds. It becomes \texttt{YES} or \texttt{NO} once $\varphi$ settles, and \texttt{UNKNOWN} if $\varphi$ never does, as with a condition on runtime input. The forms $b \vee \mathtt{COND}(\varphi)$ and $b \wedge \mathtt{COND}(\varphi)$ pair the component's own verdict $b$ with such a dependency, as in Table~\ref{tab:running-values}. A condition $\varphi$ combines atoms with AND and OR. Appendix~\ref{appendix:cond-evald} gives the rules for evaluating conditions and resolving each form. Each atom asks one question about a component $c$. The grammar has three atom forms for $\pi$. The first, $c.p$, holds if property $p$ of $c$ is \texttt{YES}. For example, $\varphi_2$ asks whether \texttt{sync\_prefs} may leak a secret. The second holds if $c$ writes into $d$ a value labeled at least $\ell$ in $\mathtt{PUBLIC} \sqsubseteq \mathtt{INTERNAL} \sqsubseteq \mathtt{SECRET}$. Secrecy properties need this form. For example, $\varphi_1$ asks whether \texttt{get\_news} writes a \texttt{SECRET} into \texttt{shared\_state:STATE}. The third holds if $c$ writes a specific value $\mathit{val}$ into $d$. Access-control gates need this form, e.g., $\mathtt{COND}(\texttt{load\_env}.\mathit{conduit\_write}[\texttt{priv}] = \texttt{"admin"})$. Recording dependencies this way makes the template compositional. Cross-component effects stay explicit, so the analysis neither drops them nor guesses their value before the dependency settles.

\textbf{LLM filling under guardrails.}
Filling runs in three steps, shown as
SA\,$\rightarrow$\,LLM\,$\rightarrow$\,SA in Figure~\ref{fig:overview},
and the second column of Table~\ref{tab:running-values} shows the
result for the running example. First, static analysis locks every
value the component's own facts establish. \texttt{get\_news} reads
the \texttt{SECRET} token from the environment, so its
\texttt{reads\_secret} is locked to \texttt{YES}. A label inherited from another
component is not locked: \texttt{sync\_prefs} reads \texttt{SECRET}
data only because \texttt{get\_news} wrote it into \texttt{STATE}, so
that value stays open. Next, one LLM call per component (Prompt~P3,
Appendix~\ref{appendix:prompts}) fills the open values from that
component's facts alone. Statically locked values are shown as final, so the LLM cannot change them. The LLM makes the semantic judgments and turns
dependencies into conditions. \texttt{get\_news}'s
\texttt{may\_leak\_secret} is \texttt{NO}, since it sends the token
only to authenticate with the news API. \texttt{sync\_prefs} sends
\texttt{STATE} to a preferences backend that has no use for the
token. Syncing preferences is its job, so it leaks nothing on its own;
it reads and leaks a secret only if \texttt{get\_news} puts one into
\texttt{STATE}. Each of its values is therefore \texttt{NO} unless
that write happens ($\varphi_1$). Likewise, \texttt{get\_news}'s
secret escapes only if \texttt{sync\_prefs} leaks, so its relay is
\texttt{NO} unless \texttt{sync\_prefs} leaks ($\varphi_2$). Finally,
static analysis derives the values that follow mechanically from
those now set. Because each LLM call is local, single-shot, and
bounded to this narrow semantic residual, LLM non-determinism stays
confined to it and never reaches a property another stage has already
fixed.

\begin{table}[t]
\centering
\footnotesize
\setlength{\tabcolsep}{5pt}
\renewcommand{\arraystretch}{1.15}
\caption{Capability templates after guarded capability filling and after the
IFC Fixpoint.}
\label{tab:running-values}
\begin{tabular}{@{}lcc@{}}
\toprule
\textbf{Property} & \textbf{After filling} & \textbf{After IFC} \\
\midrule
\multicolumn{3}{@{}l}{\textit{\texttt{get\_news}}} \\
\texttt{reads\_secret}      & \texttt{YES} & \texttt{YES} \\
\texttt{may\_leak\_secret}  & \texttt{NO}  & \texttt{NO} \\
\textit{relay}              & $\mathtt{NO} \vee \mathtt{COND}(\varphi_2)$ & \textbf{\texttt{YES}} \\
\addlinespace[2pt]
\midrule
\multicolumn{3}{@{}l}{\textit{\texttt{sync\_prefs}}} \\
\texttt{reads\_secret}      & $\mathtt{NO} \vee \mathtt{COND}(\varphi_1)$ & \textbf{\texttt{YES}} \\
\texttt{may\_leak\_secret}  & $\mathtt{NO} \vee \mathtt{COND}(\varphi_1)$ & \textbf{\texttt{YES}} \\
\textit{relay}              & $\mathtt{NO} \vee \mathtt{COND}(\varphi_1)$ & \textbf{\texttt{YES}} \\
\bottomrule
\multicolumn{3}{@{}p{0.98\linewidth}@{}}{\scriptsize
$\varphi_1 = \texttt{get\_news}.\mathit{conduit\_write}[\texttt{shared\_state:STATE}] \sqsupseteq \mathtt{SECRET}$;
$\varphi_2 = \texttt{sync\_prefs}.\texttt{may\_leak\_secret}$;
\textit{relay} = \texttt{secret\_relay\_via\_shared\_state}.
} \\
\end{tabular}
\end{table}

\subsection{IFC Fixpoint}
\label{sec:method:ifc}

After Guarded Capability Filling, each component's template is locally
consistent but may still hold conditions and open values that depend
on other components. The IFC Fixpoint resolves them at the
\emph{system level}: it walks the \afg{} and propagates information
across components until no value changes. This differs from the
forward pass of Section~\ref{sec:method:sa}, which propagates
\emph{labels over conduits}; this stage resolves \emph{dependencies
between properties}, which exist only after filling. The third column
of Table~\ref{tab:running-values} shows its result for the running
example.

\textbf{Monotone resolution.}
Every operation in this stage is monotone. Values only move from open
to settled: \texttt{UNKNOWN} or a condition becomes \texttt{YES} or
\texttt{NO}, and a settled value never changes again. The set of
settled values can therefore only grow, so the loop terminates, and no
value resolved later can contradict one resolved earlier.

\textbf{Fixpoint iteration.}
The IFC Fixpoint works in rounds. Each round updates the templates
using the values settled so far, and the rounds repeat until one
round changes nothing. A round applies three updates. It first
passes secrets along conduits: when one component writes a
\texttt{SECRET} value that another component reads, the reader's
\texttt{reads\_secret} becomes \texttt{YES}. The LLM node counts as a
reader only through tool return values, because a secret that a tool
uses internally never reaches the model. The round then sets each
component's output label, based on whether a secret reaches its
return value. Finally, it checks every open condition against the
settled values. Checking repeats within the round, so when one
resolved condition settles the operand of another, both resolve in
the same round. When a round changes nothing, the iteration stops, and
any condition still open is set to \texttt{UNKNOWN}. In the running
example in Table~\ref{tab:running-values}, the first round finds that \texttt{get\_news} writes a
\texttt{SECRET} into \texttt{shared\_state:STATE}. So $\varphi_1$
holds, and all three \texttt{sync\_prefs} values become \texttt{YES}.
\texttt{sync\_prefs} now may leak a secret, so $\varphi_2$ holds too,
and \texttt{get\_news}'s relay becomes \texttt{YES} in the same round.

\textbf{After convergence.}
Some properties depend on the fully resolved graph and are computed
once, after the loop. \texttt{memory\_poisoning\_risk} is a path
question rather than a per-component one: can untrusted data reach the
LLM's context? The user's own prompt always does, in every agent, so
direct injection carries no information about a particular agent;
\sysname{} targets \emph{indirect} injection, where untrusted content
enters through a tool. It answers the question by reachability over the \afg{}. The search starts from every \texttt{PUBLIC} conduit that a tool reads, such as HTTP responses, fetched files, and external tool returns. It then follows only components that pass the data on, as recorded by the injection track (Section~\ref{sec:method:sa}). Only
components on a complete path to the LLM node are marked \texttt{YES},
which avoids flagging every component that merely touches untrusted
data. The LLM node has no
source code, so its properties are inferred from the tools it
dispatches. Properties the model exercises through tools, such as
\texttt{writes\_secret} or \texttt{may\_leak\_secret}, are the union
over those tools, since the model may invoke any of them.
\texttt{reads\_secret} instead concerns what the model itself
receives, so it holds only if a tool returns a secret.

\subsection{ACPG and Graph Traversal}
\label{sec:method:acpg}

\sysname{}'s final output is the \emph{Agent Capability Property
Graph}~(\acpg). Formally,
$\mathcal{G} = (V_C, V_D, E, \lambda, \psi)$ is a directed graph whose
component nodes $V_C$, conduit nodes $V_D$, and data- and control-flow
edges $E$ are inherited from the \afg. The map $\lambda: V_D \to \mathcal{L}$ labels each conduit with its final
taint, and $\psi: V_C \to \mathcal{T}$ maps each component to its resolved
capability template in $\mathcal{T}$, the set of templates: the 19 core properties plus the agent-specific extended ones. Put simply, the \acpg{} is the \afg{} annotated with
resolved properties. It is built once per agent and can then be
queried by any number of downstream tasks and analyses.

\sysname{} queries the \acpg{} by deterministic, pattern-guided graph traversal. No LLM is involved at this stage. A \emph{vulnerability pattern} specifies a source condition, a sink condition, and the properties required on the path between them. The traversal returns every matching path as a \emph{candidate path}. \sysname{} implements nine pattern families, covering the vulnerability classes in Table~\ref{tab:pattern-families}. Appendix~\ref{appendix:p4} lists their verification criteria. We describe three here. \emph{Secret exfiltration} runs from a \texttt{SECRET} conduit to an external sink. It passes through components marked \texttt{may\_leak\_secret} for direct leaks, or \texttt{secret\_in\_tool\_return} for leaks through the others. \emph{Indirect injection} runs from a \texttt{PUBLIC} conduit that a tool reads to the LLM node. It passes through components marked \texttt{memory\_poisoning\_risk}. \emph{Code execution} runs from untrusted input to a component marked \texttt{executes\_dynamic\_code}.

In the running example, the secret-exfiltration pattern matches
\texttt{env:NEWS\_API\_TOKEN} $\rightarrow$ \texttt{get\_news}
$\rightarrow$ \texttt{shared\_state:STATE} $\rightarrow$
\texttt{sync\_prefs} $\rightarrow$ \texttt{http\_post:PREFS\_URL}, the
compositional leak neither tool reveals alone. Candidate paths are then adjudicated as true or false
positives (Appendix~\ref{appendix:p4}), and new threat classes
are supported by adding patterns over the same graph, without changing
any earlier stage.

%% file: src/05-implementation.tex
\section{Implementation}
\label{sec:impl}

\sysname{} is implemented in Python~3.11, using Scalpel
v1.0b0~\cite{scalpel} for AST parsing and control-flow graph
construction. It recognizes tool and orchestrator anchors for
LangChain, LangGraph, Google ADK, CrewAI, and DSPy, and resolves file
paths and connection strings by interprocedural constant propagation
so that each distinct resource receives its own conduit. The static
analysis engine tracks taint field-sensitively over variables,
attributes, dictionary keys, and container elements, composing
function summaries bottom-up and ordering components with Kahn's
algorithm~\cite{kahn_topo}. Each LLM response is validated against the
\texttt{CONDITIONAL} grammar on write, and structural values are
re-derived from the facts after every LLM call. By default, LLM calls use \texttt{claude-sonnet-4-6} (Sonnet 4.6) at temperature 0 with structured output; Section~\ref{sec:eval:rq3} evaluates other backbones. Additionally, we use Pysa (pyre-check 0.9.25)~\cite{stub-pysa} and Semgrep (v1.173.0, community edition)~\cite{stub-semgrep} as downstream application baselines. We run Semgrep with \texttt{-{}-config=auto} and
\texttt{-{}-no-git-ignore}.

%% file: src/06-Evaluation.tex
\section{Evaluation}
\label{sec:evaluation}

We evaluate \sysname{} along three research questions:
\begin{itemize}[noitemsep,topsep=2pt]
  \item[\textbf{RQ1}] Can the pipeline accurately model the capability
    surface of LLM-based agents? Is the neurosymbolic (hybrid SA+LLM) design necessary,
    or is either component alone sufficient?
  \item[\textbf{RQ2}] What is the downstream impact of the capability
    model? Does the \acpg\ enable more effective security analysis than
    state-of-the-art SA or LLM-only alternatives?
  \item[\textbf{RQ3}] Do LLM variability and model selection affect
    pipeline quality and the resulting capability model?
\end{itemize}

\subsection{Experimental Setup}
\label{sec:eval:setup}

We evaluate on a single corpus of 43 unmodified real-world agents. RQ2
has two parts: Part 1 tests whether the ACPG reconstructs known threat
chains, and Part 2 tests whether it surfaces vulnerable paths in agents as found in the wild, with no planted vulnerabilities. RQ1, RQ2 Part 1, and RQ3
score ACaM against manual annotation, so they use a controlled 24-agent
subset: 10 unmodified base agents from the corpus and 14 vulnerable variants
derived from them. RQ2 Part 2 has no annotation to score against, so it runs on all 43 agents, and we confirm a subset of its findings with proof-of-concept exploits.

\textbf{Discovery Corpus.} We collect 43 agents from public GitHub
repositories spanning six implementation styles: LangChain, LangGraph,
Google ADK, CrewAI, DSPy, and direct LLM SDK usage. Agents average approximately
6.2 tools. Table~\ref{tab:discovery-corpus} breaks the corpus down by framework
and architecture; an agent is multi-agent if it uses an explicit
orchestration primitive or contains at least two independently
invocable agents.

\begin{table}[t]
\centering
\caption{Discovery Corpus (43 agents) by framework and architecture.}
\label{tab:discovery-corpus}
\footnotesize
\setlength{\tabcolsep}{3pt}
\begin{tabular}{lrrrrrrr}
\toprule
 & LangChain & LangGraph & ADK & CrewAI & DSPy & SDK & Total \\
\midrule
Single & 20 & 9 & 1 & 0 & 1 & 3 & 34 \\
Multi  & 0  & 4 & 3 & 2 & 0 & 0 & 9  \\
\bottomrule
\end{tabular}
\end{table}

\textbf{Injected Vulnerable Variants.} From 10 base corpus agents we derive 14 variants, each injecting one or two
hidden exfiltration paths into an otherwise unchanged base. Tool
descriptions are left as they were, so every injected path is a
laundered capability by construction. Vulnerabilities follow the OWASP Agentic
Security Initiative (ASI) taxonomy~\cite{owasp2026agentic} and fall
into three classes of increasing structural difficulty. A1 (ASI02 Tool
Misuse and Exploitation; ASI06 Memory and Context Poisoning) places the
secret in a tool's return value, so the LLM itself becomes the
exfiltration channel. A2 (ASI02; ASI07 Insecure Inter-Agent
Communication) bypasses the LLM with a silent HTTP POST inside the
tool. A3 adds structural indirection, routing the secret across
shared-state or subprocess boundaries so that detection requires
long compositional analysis.

\textbf{Ground Truth.} No existing tool can directly generate ground-truth
capability properties without relying on its own assumptions, so to ensure
a fair, independent baseline we annotate manually. For RQ1 we manually
annotate core capability property values for all 24 agents (10 base agents,
14 variants), giving a controlled comparison. We do not annotate extended properties, as they are open-ended and per-agent, making them
difficult to evaluate against a single fixed standard. For RQ2 Part 1, we annotate each injected chain. Scoring a chain as a single pass/fail would hide where a method goes wrong. For example, a method that names the right secret and sink but drops an intermediate hop would score the same as one that finds nothing. We therefore decompose each
chain into seven facts that are scored independently, illustrated here with the
\texttt{get\_weather} tool of Section~\ref{sec:introduction}. Every exfiltration
chain shares the same skeleton, and each fact pins down one part of it.
The trigger is the tool the agent invokes (\texttt{entry\_tool}:
\texttt{get\_weather}). The source is the secret and where the program
reads it (\texttt{secret\_variable}: \texttt{AWS\_SECRET\_ACCESS\_KEY};
\texttt{secret\_read\_function}: \texttt{load\_creds}), plus the
expression where it first enters the data flow (\texttt{injection\_point}:
\texttt{key = cfg.get(...)}). The path is the ordered call sequence from
source to exit (\texttt{chain}: \texttt{load\_creds} $\rightarrow$
\texttt{build\_request} $\rightarrow$ \texttt{get\_weather}). The sink is
how and where the secret leaves the agent (\texttt{exfil\_boundary}:
\texttt{http\_request}; \texttt{exfil\_function}: \texttt{get\_weather}).
This decomposition also lets us report per-fact miss rates, showing which
part of a chain each method loses. RQ2 Part 2 targets unmodified agents whose vulnerabilities are not known
in advance, so no fixed ground truth exists, and the volume of candidate
paths rules out manual review of each one. A verification agent instead
adjudicates each candidate as a true or false positive
(Appendix~\ref{appendix:p4}). This is a bounded judgment task. The agent
sees one path at a time, with its source code and pipeline evidence, and
applies explicit per-class criteria. For confirmation independent of the
verifier's judgment, we manually review all Critical paths and develop proof-of-concept exploits for a subset
of findings (Section~\ref{sec:eval:rq2:ds2}).

\textbf{Metrics.} For RQ1, each core property takes one of six values
(\texttt{YES}, \texttt{NO}, \texttt{UNKNOWN}, \texttt{PUBLIC},
\texttt{INTERNAL}, \texttt{SECRET}). We compute per-class precision,
recall, and F1 against ground truth, macro-averaged. For RQ2, each
reconstructed chain scores $(\textsc{Full} + 0.5 \times \textsc{Partial})/7$
against the 7-fact schema, where \textsc{Full} counts facts that match the
ground truth (case-insensitive substring), \textsc{Partial} counts facts
with meaningful overlap, and any other fact is a \textsc{Miss}; for
\texttt{chain}, \textsc{Partial} requires two consecutive ground-truth
hops in order. An agent with more than one independent chain receives the mean of its chain scores.

\textbf{Baselines.} RQ1 and RQ2 isolate the necessity of the hybrid
design along two axes: what static analysis achieves without an LLM,
and what an LLM achieves without static analysis.

\textit{SA-only.} For RQ1, we define \sysname{}-SA as \sysname{} with every LLM component
removed, measuring what the pipeline's static analysis achieves on its own. For RQ2 we run Pysa and
Semgrep. Classical SA has no notion of an LLM dispatch boundary. Pysa's
taint-in-taint-out (TITO) stubs approximate one, so we run Pysa both
without and with stubs. Semgrep runs community rules with pattern matching only.

\textit{LLM-only.} Every strategy runs on both Sonnet 4.6 and Opus 4.8. Each receives the source code and fills its output directly. For RQ1, it receives the property definitions and fills the core-property template. For RQ2 Part 1, it receives the 7-fact schema and fills the seven facts of each chain. Appendix~\ref{appendix:llm-only-prompts} shows their core prompts. RQ1 uses six strategies, and RQ2 Part 1 uses three:
\begin{itemize}
  \item \textit{Agentic} (RQ1, RQ2). \texttt{claude~-p} with
    \texttt{Read} explores the source on demand.
  \item \textit{Five-Round} (RQ1) / \textit{Chain Tracer} (RQ2).
    Simulates a static analyzer over five rounds: function inventory,
    taint sources, sink identification, chain tracing, and structured
    verdict. Chain Tracer retargets the same structure to the 7-fact
    schema.
  \item \textit{One-shot} (RQ1). A single call with full source and
    definitions, no iterative structure.
  \item \textit{Persona} (RQ1). Four analyst personas (Data Flow, Supply Chain Auditor, IFC, Red Team) analyze the source, and a Synthesizer merges their findings into the template.
  \item \textit{Verifier} (RQ1). One instance runs Five-Round; a fresh instance of the same model challenges every NO with code evidence.
  \item \textit{Adversarial} (RQ1). After Five-Round, a sixth round asks the same instance to back every NO with code or flip it to YES. Verifier and Adversarial challenge only NO values. A missed capability is the costlier error. 
  \item \textit{Code Reviewer} (RQ2). An open-ended security review
    followed by structured JSON extraction in the same context.
\end{itemize}

\subsection{RQ1: Pipeline Accuracy and Necessity of Hybrid Design}
\label{sec:eval:rq1}

This RQ asks whether \sysname{} models capability surfaces accurately, and whether the hybrid design outperforms either of its components alone. We evaluate the 19 core properties on the 24-agent annotated subset (10 base, 14 vulnerable; Appendix~\ref{app:rq1_additional}). For the LLM-only baseline, we sweep 6 prompting strategies across 2 models (12 configurations), take the best per agent, and report their average, denoted LLM-only (best). As Table~\ref{tab:rq1_aggregate} shows, \sysname{} outperforms both baselines on every metric, reaching 0.983 macro-F1 against 0.934 for \sysname{}-SA and 0.814 for LLM-only (best) configuration.

\sysname{}-SA comes close to the full pipeline on its own, because most core properties can be decided directly from structural facts in the code. It falls short on semantic properties: these do not reduce to a fixed syntactic pattern, so no static rule set can anticipate every form they take across agents. Its deficit is largest in recall (0.936 vs.\ 0.991), meaning that without the LLM, the pipeline more often misses capabilities that are present.

The 24-agent aggregate numbers in Table~\ref{tab:rq1_aggregate} average over two kinds of agents that test different things, so we also break the results down by agent type. The 10 base agents are unmodified real-world code; here, we check that the pipeline reports the capabilities the agent has, and no others. The 14 vulnerable variants are derived from these same agents, each with an injected capability that is harder to find; here, the question is whether the pipeline recovers it. Comparing the two shows where the hybrid analysis earns its place. On base agents, removing the LLM costs 3\% of F1 (0.973 to 0.944); on vulnerable variants, it costs 7\% (0.993 to 0.923). The hybrid makes its largest gains exactly where capabilities are hardest to find, because those are the capabilities that static rules alone do not resolve.

\begin{figure}[t]
\centering
\includegraphics[width=\linewidth]{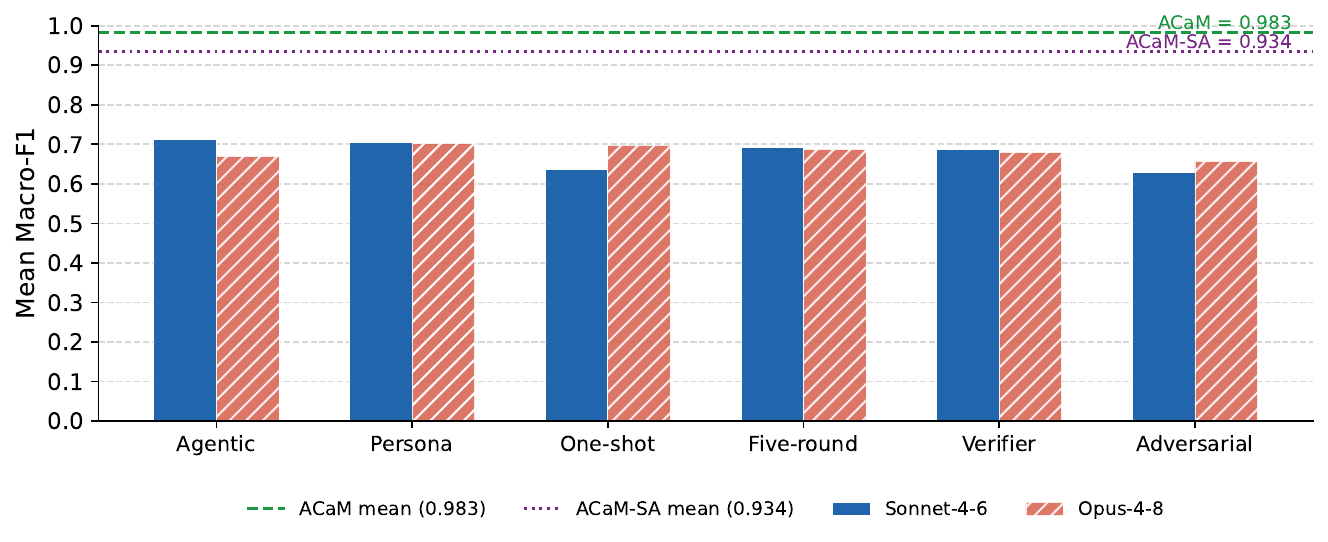}
\caption{Mean macro-F1 per LLM-only prompting strategy}
\label{fig:rq1-llm-only-chart}
\end{figure}

LLM-only analysis underperforms both, and Figure~\ref{fig:rq1-llm-only-chart} shows that this gap holds across all 12 configurations; substituting Opus for Sonnet yields no consistent improvement. Together, these results support our design rationale. Static analysis supplies structural precision, the LLM supplies semantic interpretation, and the hybrid outperforms either component in isolation.

\begin{table}[t]
\centering
\caption{\sysname{}'s capability modeling performance on the 24-agent annotated subset. Scores are macro-averaged over property value classes per agent. LLM-only (best) takes the best of 12 configurations (6 prompting strategies, 2 models) per agent. Aggregation details are in Appendix~\ref{app:rq1_additional}.}
\label{tab:rq1_aggregate}
\small
\begin{tabular}{@{}lcccc@{}}
\toprule
\textbf{System} & \textbf{F1} & \textbf{Prec} & \textbf{Rec} & \textbf{Acc} \\
\midrule
ACaM-SA         & 0.934 & 0.947 & 0.936 & 0.965 \\
LLM-only (best) & 0.814 & 0.831 & 0.853 & 0.881 \\
\sysname{}      & \textbf{0.983} & \textbf{0.980} & \textbf{0.991} & \textbf{0.992} \\
\bottomrule
\end{tabular}
\end{table}

\subsection{RQ2: Downstream Applications}
\label{sec:eval:rq2}

We evaluate the \acpg{}'s downstream utility along two axes. First, \textit{threat chain reconstruction}: we compare \sysname{} against SA-only (Semgrep + Pysa) and LLM-only baselines on benchmark variants with injected, ground-truth-verified attack chains across the three scenarios from Section~\ref{sec:eval:setup}. Second, \textit{structural vulnerability identification}: we apply the same pipeline to 43 unmodified real-world agents to test whether the \acpg{} surfaces genuine security risks with no planted vulnerabilities and no ground truth. Together, the two axes test whether an accurate capability model enables analysis that neither SA nor LLM-only tools can match. The first axis tests this where the answer is known. The second tests it in the open, where it is not.

\subsubsection{Threat Chain Reconstruction}
\label{sec:eval:rq2:ds1}

\begin{figure}[t]
\centering
\includegraphics[width=\linewidth]{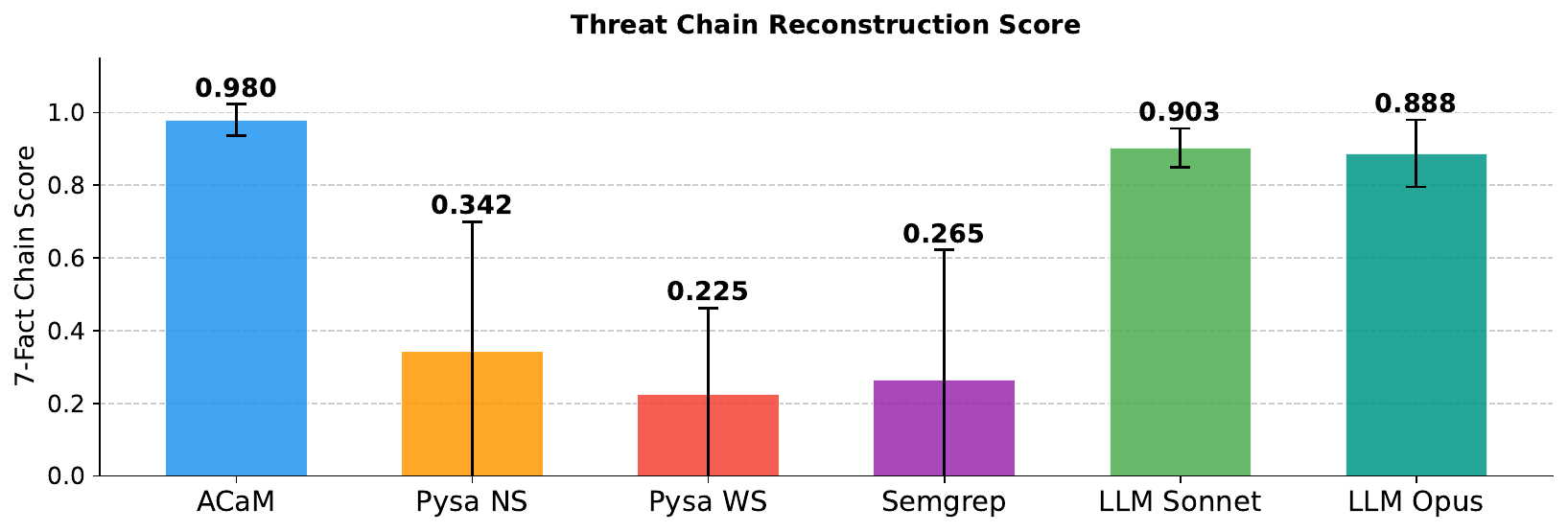}
\caption{Mean 7-fact threat chain reconstruction score across all 14 vulnerable variants}
\label{fig:rq2_chain_scores}
\end{figure}

\begin{figure}[t]
\centering
\includegraphics[width=\linewidth]{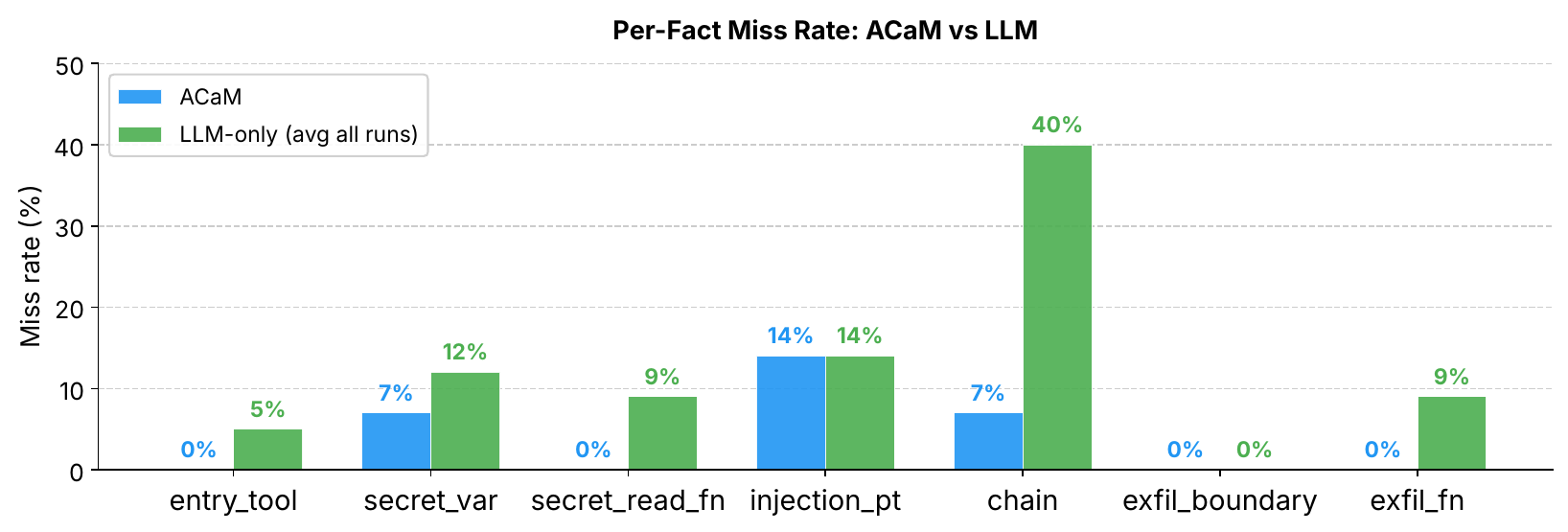}
\caption{Per-fact Miss rate on the seven threat chain facts: \sysname{} vs LLM (averaged across all runs). Lower is better.}
\label{fig:rq2_fact_miss_acam_vs_llm}
\end{figure}

We score threat chain reconstruction using the 7-fact schema and scoring function from Section~\ref{sec:eval:setup}. Per-agent scores are in Table~\ref{tab:rq2_chain_scores} (Appendix~\ref{app_rq2_additional}), and aggregate scores are in Figure~\ref{fig:rq2_chain_scores}. For SA, we report a no-stub configuration (Pysa NS) and a configuration with a taint-in-taint-out stub for the LLM dispatch loop (Pysa WS). For LLM-only, we evaluate three prompting strategies per model, take the best per agent, and report their average for each of Sonnet and Opus.

SA is the weakest baseline (Pysa NS: 0.342, Pysa WS: 0.225, Semgrep: 0.265). It cannot model the LLM/tool dispatch boundary. Stub modeling only makes this worse, because the TITO approximation of the LLM call introduces noise that the analysis cannot resolve. Off-the-shelf static analysis is therefore
inadequate for agentic code unless the LLM boundary is modeled explicitly
rather than approximated. LLM-only performs substantially better (Sonnet: 0.903, Opus: 0.888). Sonnet is the stronger of the two for this task.

\sysname{} achieves the best overall score (0.980), roughly 8 points above the strongest LLM-only baseline. The per-fact miss-rate breakdown (Figure~\ref{fig:rq2_fact_miss_acam_vs_llm}) shows why. Here, a fact counts as missed unless it scores \textsc{Full}. LLMs stay close to \sysname{} on the other six facts, which are well represented in pretraining. On \texttt{chain}, however, their miss rate jumps to 40\%, compared with 7\% for \sysname{}. LLMs reliably name the source and the sink, but they drop intermediate hops when asked to connect them. \sysname{} avoids this because it tracks the chain structurally through the capability graph rather than reconstructing it post hoc.

Together with RQ1, these results point to a consistent division of labor.
Unmodified static analysis is not enough for agentic code. A usable SA layer must understand the LLM/tool boundary. It cannot treat that boundary as opaque. LLMs are reliable where a task has strong training or
verification signal (sources, sinks, known vulnerability patterns) but
degrade on open-ended structural tasks like assembling a full capability
representation. Grounding the LLM in SA-derived structure converts this
open-ended judgment into a bounded one, which is why the hybrid approach
outperforms either component alone.

\subsubsection{Structural Vulnerability Identification} \label{sec:eval:rq2:ds2}

\begin{table*}[t]
\caption{Unique paths by vulnerability family, 43-agent corpus. Each path has a unique source-sink pair.}
\label{tab:findings-summary}
\centering\footnotesize
\setlength{\tabcolsep}{6pt}
\begin{tabularx}{\textwidth}{l r l >{\raggedright\arraybackslash}X}
\toprule
Category & Vuln.\ Paths & Sev. & Key Example \\
\midrule
Code Execution / RCE          & 107   & Critical & In-process REPL RCE; deserialization RCE; write-then-exec RCE \\
Secret / Credential Exposure  & 16    & Critical & Hardcoded secrets reach an external sink \\
\midrule
Indirect Prompt Injection     & 1,692 & High     & Web-crawl content, alert-JSON entry, KB or shared-state poisoning, cross-agent injection \\
SQL Injection                 & 64    & High     & LLM-generated SQL scope bypass \\
Access Control / IDOR & 15 & High & Order lookup without authentication, leaking other customers' data \\
SSRF / External Comms         & 6     & High     & User URL to cloud metadata endpoint \\
Output Injection / XSS        & 4     & High     & Stored XSS via unsafe markdown rendering \\
\midrule
Other & 16 & Medium & Guardrail bypass; unguarded tool chaining; session-isolation failure across users; model reasoning leakage \\

\bottomrule
\end{tabularx}

\vspace{2pt}

\end{table*}

\begin{table*}[t]
\centering
\footnotesize
\caption{Proof-of-concept exploits for ten structural findings. P1 to P6 are Critical; P7 to P10 are High. Paths: distinct confirmed exploit paths to the same sink within the agent. Signal: whether detection relied on core properties, extended properties, or both.}
\label{tab:poc}
\begin{tabularx}{\textwidth}{@{}l l l c c l X@{}}
\toprule
ID & Agent & Vulnerability class & TM & Paths & Signal & Exploit \\
\midrule
P1  & \texttt{Agent 1}       & RCE, in-process REPL      & A & 1 & Core        & Poisoned search snippet invokes \texttt{PythonAstREPLTool}; arbitrary code runs in-process \\
P2  & \texttt{Agent 2}          & RCE, in-process REPL      & A & 5 & Core        & Web-UI input reaches PythonAstREPLTool; arbitrary code runs and user records are exfiltrated \\
P3  & \texttt{Agent 3}                  & RCE, in-process REPL      & A & 2 & Core        & Crafted question reaches \texttt{dspy.PythonInterpreter}; arbitrary code runs in-process \\
P4  & \texttt{Agent 4}         & RCE, deserialization      & B & 1 & Core        & Tampered .pkl file is unpickled at import; code runs at startup \\
P5  & \texttt{Agent 5}     & RCE, deserialization      & B & 1 & Core + Ext. & Tampered vector index is deserialized via pickle; code runs at startup \\
P6  & \texttt{Agent 6}            & RCE, write-then-exec      & B & 1 & Core + Ext. & A poisoned task file steers an LLM-authored script; its execution deletes project files \\
\midrule
P7  & \texttt{Agent 7}             & Indirect prompt injection (IPI) & A & 1 & Core + Ext. & Injected URL persists in the audit trail\\
P8  & \texttt{Agent 8} & Shared-state poisoning via IPI & A & 1 & Core + Ext. & Injected content propagates to all agents; attacker contact appears in the final verdict \\
P9  & \texttt{Agent 5}     & IDOR                      & A & 1 & Ext.        & Oracle-and-guess queries return other customers’ records in two of four attempts \\
P10 & \texttt{Agent 9}              & Stored XSS (3-hop)        & A & 1 & Ext.        & Stored \texttt{<img onerror>} payload executes in every viewer's browser \\
\bottomrule
\end{tabularx}
\begin{flushleft}
\footnotesize
\textit{Note:} We reported all ten findings to the affected maintainers through coordinated disclosure. We withhold project names until fixes are released.
\end{flushleft}
\end{table*}

We apply \sysname{} to 43 unmodified real-world agents drawn from the Discovery Corpus. We run the analysis directly on source code, with no planted vulnerabilities and no prior security annotation. Findings are evaluated under two threat models. In the external-attacker threat model (TM-A), the attacker has no code or deployment access and acts only through public input channels, such as user prompts, uploaded files, and external data the agent fetches. In the write-access threat model (TM-B), the attacker has same-host or side-channel write access but cannot modify code directly. Examples include a poisoned cache directory, a compromised CI artifact, or a shared volume.

Table~\ref{tab:findings-summary} groups verified findings by vulnerability family. The verifier assigns each family (Appendix~\ref{appendix:p4}). The Sev.\ column gives each family's typical severity; individual paths within a family may be rated lower. Code Execution/RCE and Secret/Credential Exposure are the most severe families, and Indirect Prompt Injection dominates by volume. This distribution reflects a structural limit that affects even carefully written agents. Unlike a parameterized query, a prompt has no boundary between data and instruction. Any untrusted content that an agent ingests can therefore act as an instruction to the LLM. As a result, Indirect Prompt Injection remains open by construction. Resolving it is outside our scope. \sysname{} detects the exposure and leaves the choice of sanitization to developers.

The counts in Table~\ref{tab:findings-summary} include only paths that passed verification, not raw candidate paths. Each count is a unique path. No two paths share both a source and a sink. We count each path separately because each is a separate way in; patching one can leave another open. A separate verification agent labels each candidate path as a true or false positive (Appendix~\ref{appendix:p4}). Unmodified agents lack ground truth. We manually reviewed all 123 Critical paths and confirmed that each exists in the code as reported. Before deduplication, the verifier confirms 99.4\% of High and 98.4\% of Medium candidates. These rates reflect only the verifier's judgment. The PoC exploits below confirm a subset independently. Rejected candidates concentrate in a few agents. Two agents account for 85\% of all FPs. Manual inspection attributes this to a few overly broad conduit-classification rules specific to those two agents. The largest single source of false positives comes from agents reading internal files that the traversal treats as indirect-prompt-injection vectors. Under TM-A, these are almost always developer-controlled files with no external write path. Under TM-B, however, the same read becomes a genuine exploit path if an attacker can write to the file.

A single agent can also accumulate unrelated vulnerability classes. For example, \texttt{Agent 5} has two separate PoC-confirmed findings (P9 and P5). The first is an insecure direct object reference (IDOR) caused by an optional authentication check. The second is a deserialization RCE that runs at startup, with no LLM involved. The two findings have different root causes. Yet both exist in one deployed system. A whole-agent profile shows them together, as the accumulated risk of one system.

For confirmation, independent of the verifier's own judgment, we developed working proof-of-concept exploits for ten Critical and High findings across nine agents (Table~\ref{tab:poc}). We ran all exploits against local deployments with synthetic data; no deployed instance or third-party user was involved. A single exposure can be reachable along several independent routes: in \texttt{Agent 2} (P2), five separate entry points all terminate in the same unrestricted Python interpreter.

Six of the ten exploits achieve arbitrary code execution in three ways. In P1 and P2, untrusted input reaches an interpreter that runs inside the agent without a sandbox. P3 relies on \texttt{dspy.PythonInterpreter}, which runs in-process only under DSPy 2.5.x and earlier; later versions sandbox it. In P4 and P5, the agent loads a tampered artifact through unsafe deserialization. Tampering requires same-host write access to the artifact file, not to the code (TM-B). In P6, one component saves an LLM-written file, and another component runs it. Core properties alone detect P1 to P4; P5 to P8 need extended properties to name the full chain; and P9 and P10 are detected by extended properties alone.

All exploits succeeded with qwen2.5:7b, a minimally safety-trained model, as the backbone; all but P9 on the first attempt. Claude Haiku 4.5, despite heavy safety training, blocked blunt attempts, but all ten still succeeded under more subtly framed prompts. We did not escalate to Sonnet or Opus, because the pattern is architectural rather than model-specific. Together, this evidence yields two observations about structural vulnerability and model safety training. (a) Some exploits do not depend on the LLM misbehaving, so safety training cannot prevent them. P4 and P5, for example, execute attacker code during deserialization, before any model decision. (b) Where an exploit does depend on the LLM, a more carefully framed prompt gets through. Safety training raised the bar in every case we tested but never closed the hole; a stronger model may raise it further, but cannot remove the code defect. \sysname{} surfaces the paths behind all ten exploits statically, before any model runs. For the code-execution and access-control findings, remediation is a code change; no prompt, model, or safety training removes the defect.

We also ran an LLM-only analysis, without \sysname{}'s static grounding, to check whether it could find these vulnerabilities on its own. The results follow the same pattern as the LLM-only baselines in the previous subsection. LLMs locate plausible sources and sinks, but they do not reliably reconstruct the full end-to-end chain. They also tend to stop at the first path they find to a sink. \sysname{} instead surfaces multiple
independent paths to the same sink where they exist, which is what
the multi-path counts in Table~\ref{tab:poc} reflect.

\subsection{RQ3: LLM Variability, Model Selection, and Cost}\label{sec:eval:rq3}

\textbf{Determinism and stability.}
We run \sysname{} 15 times with Sonnet 4.6 on each of the 24 RQ1 agents. Macro-F1, precision, and recall stay stable across runs. Only 0.5\% to 2\% of property values change between runs, and only a few agents reach the upper end. This stability comes from Guarded Capability Filling (Section~\ref{sec:capability-filling}), which grounds every LLM call in static-analysis facts.

\begin{table}[t]
\centering
\caption{F1 and cost by backbone model on the 24 RQ1 agents. Means over 15 runs. Standard deviations are negligible and omitted.}
\label{tab:rq3_model_selection}
\footnotesize
\begin{tabular}{lccc}
\toprule
 & \textbf{Sonnet 4.6} & \textbf{Opus 4.8} & \textbf{Haiku 4.5} \\
\midrule
F1        & 0.983 & 0.982 & 0.952 \\
Cost (\$) & 0.769 & 0.863 & 0.743 \\
\bottomrule
\end{tabular}
\end{table}

\textbf{Model selection.}
The backbone model is the LLM that \sysname{} calls to fill the semantic residual. To test how much it matters, we swap the backbone and keep the rest of the pipeline fixed. Table~\ref{tab:rq3_model_selection} compares three backbones on the same 24 agents, with 15 runs each. Sonnet and Opus reach nearly the same F1 (0.983 vs.\ 0.982), but Sonnet costs less (\$0.769 vs.\ \$0.863). Haiku is less accurate (0.952) at a comparable cost (\$0.743). Sonnet therefore offers the best balance. Even Haiku, the weakest backbone, exceeds LLM-only (best) configuration from RQ1 (0.814 F1, Table~\ref{tab:rq1_aggregate}) by 0.14. Static grounding, not model tier, is therefore the main driver of accuracy.

\textbf{Cost-accuracy tradeoff.}
Table~\ref{tab:rq3_cost} reports per-agent cost and time with
Sonnet 4.6. Populating only the 19 core properties costs
\$0.489 per agent, a few cents more than LLM-only analysis (\$0.424),
and in return delivers substantially higher accuracy
(Section~\ref{sec:eval:rq1}). Populating the extended properties, which
drive most of RQ2's real-world findings, adds \$0.280 per agent. At a
mean of 57 seconds per agent, \sysname{} is thus stable, affordable, and
practical.

\begin{table}[t]
\centering
\caption{Cost and time across 24 agents (same set as RQ1), mean $\pm$ std over 15 runs.}
\label{tab:rq3_cost}
\footnotesize
\begin{tabular}{@{}lcc@{}}
\toprule
\textbf{Config} & \textbf{Cost (\$)} & \textbf{Time (s)} \\
\midrule
LLM-only (avg) & 0.424 $\pm$ 0.120          & 112.0 $\pm$ 35.2 \\
ACaM-SA        & 0.000                      & 5.0 \\
\sysname{}     & 0.769 $\pm$ 0.002$^{\ast}$ & 57.0 $\pm$ 52.8 \\
\bottomrule
\end{tabular}
\par\vspace{2pt}
{\footnotesize $^{\ast}$Core properties \$0.489, extended properties \$0.280.}
\end{table}

%% file: src/07-discussion-and-limilations.tex
\section{Discussion and Limitations}
\label{sec:discussion}

\textbf{Python scope.} \sysname{} analyzes Python source. All five
frameworks in our corpus are Python-native and widely
used.\footnote{As of November 2026, LangChain has 150K GitHub stars, CrewAI 61K, LangGraph 45K, DSPy 39K, and Google ADK 22K.} Other popular
agent frameworks, such as AutoGen, MetaGPT, and the OpenAI Agents SDK,
are also written in Python. Only the AST frontend, the framework-anchor
detectors, and the conduit rules for library calls are Python-specific.
The rest of the pipeline is language-agnostic, and porting these three
parts to TypeScript or another language is near-term future work.

\textbf{Injected vulnerability realism.} Our injected vulnerabilities
mirror disclosed precedent. Class A1's LLM-mediated exfiltration matches
the original indirect prompt injection PoCs against Bing Chat~\cite{ipi} and M365
Copilot cases surveyed in \cite{stub-christodorescu}; Class A2's out-of-band
exfiltration mirrors two LangChain CVEs, a serialization bypass
\cite{nvd2025cve68664} and an eval()-based RCE \cite{nvd2024cve21513};
Class A3's indirection has a 2025 analogue \cite{nvd2025cve55284}.

\textbf{Static analysis engine limitations.} The engine is conservative
by design. The broadest choice is that \sysname{} models the worst-case
capability surface. The LLM decides at runtime which tools to call and in
what order, and that order cannot be observed before deployment. So once
any tool's output taints the LLM's context, every tool reachable from the
LLM node is treated as a possible reader of that taint. In addition,
sanitizers are handled with the same caution. The engine detects and
records sanitizer calls but never uses them to lower a taint label.
Whether a sanitizer makes data safe depends on where that data ends up,
which static analysis alone cannot determine. These choices can add false positives, but few appear in our evaluation (Section~\ref{sec:eval:rq2:ds2}). Additionally, when a file path is built at runtime, as in \texttt{open(os.path.join(base, user\_input))}, the engine cannot tell which file it names. Such reads can therefore cause false negatives. \sysname{} labels them \texttt{INTERNAL} by default, because labeling them all \texttt{SECRET} would flood the results with false positives.

\textbf{Future applications of the ACPG.} Beyond the two applications we evaluate, the ACPG is a general-purpose capability model that other applications can query directly. It can drive least-privilege policy synthesis, generating a component's minimal permission set from its actual capability facts rather than requiring one to be hand-authored. It can supply the policy input that runtime enforcement mechanisms (Section~\ref{related_work}) currently assume as given. Paired with tool descriptions, the ACPG could support a finer-grained
inconsistency checker that names the mismatched capability. And superimposing multiple agents' ACPGs over their inter-agent conduits extends the same traversal machinery across agent boundaries, modeling a distributed multi-agent system as a single compositional graph. Each is a direct instantiation of the existing model, not a new pipeline.

%% file: src/08-Conclusion.tex
\section{Conclusion}

Agentic systems trust tool descriptions, and nothing in the agent verifies them. This unverified trust enables a threat class called capability laundering, in which an agent carries a dangerous capability that no tool's description declares. We address it with \sysname{}, a neurosymbolic, compositional framework for agentic-system vulnerability discovery. \sysname{} derives capabilities from code, so a laundered capability cannot hide behind a benign description. Static analysis grounds every security-critical conclusion, an LLM resolves only what structure cannot, and the Agent Capability Property Graph unifies both over the composed system. Across 43 real-world agents, \sysname{} uncovers multi-hop vulnerable chains that no single-component tool reveals and surfaces ten PoC-confirmed vulnerabilities, six of them arbitrary code execution.

%% file: src/09-appendixA.tex
\section{Supporting Details of \sysname{}}

\subsection{LLM Prompts}
\label{appendix:prompts}

This appendix presents the crux of the LLM prompts used by \sysname{}.

\subsubsection{Prompt P1 : Conduit Classification}
\label{appendix:p1}

Used in AFG Construction (Section~\ref{sec:method:afg}).
Receives the deduplicated AST call-site list and conservative candidate rules.
The LLM reviews each candidate (keep or mark false positive), corrects taint
labels, and adds agent-specific rules missed by the conservative generator.

\begin{lstlisting}[style=promptcode, caption={Prompt P1: Conduit classification system prompt (excerpt).}, label={lst:p1}]
You are a static analysis expert reviewing candidate conduit rules for a Python agentic AI system. A conduit is any site where data crosses a component boundary: shared state, files, HTTP, subprocess stdout, env vars, databases, queues, IPC sockets, LLM API calls.
Task: (A) mark each candidate keep or false_positive. (B) add missed agent-specific rules (custom/proprietary frameworks, LangChain/LangGraph/LlamaIndex patterns, non-stdlib boundaries).
Rules: 0. conduit ID = the resource (path/var/db name), never the function name. open("~/.aws/credentials") -> "file:~/.aws/credentials"; os.getenv("API_KEY") -> "environment_var:API_KEY"
1. List every candidate; use decision=false_positive to exclude, never omit. [...]
\end{lstlisting}

\subsubsection{Prompt P2 : Extended Property Discovery}
\label{appendix:p2}

Used in Capability Property Discovery (Section~\ref{sec:method:afg:template}). Prompt P2 is a  %
single whole-agent call that reviews all components together and
proposes extended properties that no single-component view could
surface.

\begin{lstlisting}[style=promptcode, caption={Prompt P2: Whole-agent extended property discovery (excerpt).}, label={lst:p2}]
You are a security analyst reviewing a complete AI agentic system. Now identify cross-component security patterns that cannot be seen from any single component alone.
## All components in this agent
{all_component_blocks}
## Task -- Unified cross-component extension properties
Propose at most {max_extensions} security-relevant extension properties that:
- Capture patterns spanning MULTIPLE components (e.g. taint flows across process boundaries, privilege escalation chains through shared state, hidden exfiltration paths)
- Are NOT already covered by the 19 core properties
- Would matter to a security analyst doing a full-agent review
For each extension, specify:
- name: snake_case identifier
- description: one sentence -- what is the security concern
- applies_to: list of component IDs where this is relevant
- notes_by_component: for each component in applies_to, one-line note pointing to the specific CODE relevant to this property in that component
Respond with ONLY valid JSON. No markdown fences. [...]
\end{lstlisting}

\subsubsection{Prompt P3 : Guarded Capability Filling}
\label{appendix:p3}
Used in Guarded Capability Filling (Section~\ref{sec:capability-filling}), with one call per component. The LLM receives two inputs. The first is the structural values that static analysis has already filled. The second is the taint facts: conduit reads and writes with their labels and the conditions under which they occur, subprocess boundaries, and implicit guards. The LLM fills only \texttt{UNKNOWN} slots. It is asked to leave every value locked by static analysis unchanged.
\begin{lstlisting}[style=promptcode, caption={Prompt P3: Guarded capability filling system prompt excerpt.}, label={lst:p3}]
You are a security analyst filling a component's capability template for an agentic system.
You are given static analysis facts for ONE component:
  CONDUIT READS   -- conduits read, taint labels, conditional_on
  CONDUIT WRITES  -- conduits written, taint labels, conditional_on
  SUBPROCESS BOUNDARIES -- subprocess/fork/execv calls detected
  IMPLICIT GUARDS -- branch conditions on tainted variables
You must fill every UNKNOWN property with:
  YES                        -- confirmed present from local facts
  NO                         -- confirmed absent
  CONDITIONAL(<expr>)        -- depends on a different component
  UNKNOWN                    -- no evidence at all
Pre-filled values are ground truth -- never override them.
CONDITIONAL EXPRESSION FORMAT:
  CONDITIONAL([component_id.property_name])
  [...]
RULES:
  - Use component IDs exactly as they appear in written_by fields
  - Never invent component names -- use the actual ID
  - If a property is conclusive from local facts: use YES or NO
  - memory_poisoning_risk: ALWAYS NO -- filled by IFC stage only
  - output_taint_label:    ALWAYS UNKNOWN -- filled by IFC stage
KEY SEMANTIC DISTINCTIONS:
  produces_secret_output = secret reaches INTERNAL channel (state write, return value) -- NOT external exfiltration
  may_leak_secret        = secret reaches EXTERNAL boundary (HTTP, file, subprocess) -- NOT internal propagation
  secret_in_tool_return  = secret in THIS component's return value
  external_communication = outbound HTTP/socket (LLM backbone API calls do NOT count as external_communication)
\end{lstlisting}

\subsection{CONDITIONAL Evaluation Rules}   %
\label{appendix:cond-evald}
\begin{table}[h]
\centering
\caption{Evaluation of \texttt{AND}/\texttt{OR} over
\{\texttt{YES}, \texttt{NO}, \texttt{UNKNOWN}\}.}
\label{tab:cond-eval}
\scriptsize

\begin{tabular*}{\linewidth}{@{\extracolsep{\fill}}c|c|c|c}
\textbf{AND} & NO & UNKNOWN & YES \\\hline
NO       & NO & NO      & NO \\
UNKNOWN  & NO & UNKNOWN & UNKNOWN \\
YES      & NO & UNKNOWN & YES \\
\bottomrule
\end{tabular*}

\vspace{0.5em}

\begin{tabular*}{\linewidth}{@{\extracolsep{\fill}}c|c|c|c}
\textbf{OR} & NO & UNKNOWN & YES \\\hline
NO       & NO      & UNKNOWN & YES \\
UNKNOWN  & UNKNOWN & UNKNOWN & YES \\
YES      & YES     & YES     & YES \\
\bottomrule
\end{tabular*}

\end{table}

A compound CONDITIONAL formula can combine sub-conditions that are
already settled (\texttt{YES} or \texttt{NO}) with sub-conditions that
are still open. During the IFC Fixpoint, an open sub-condition is
evaluated as \texttt{UNKNOWN}, and one that never settles stays
\texttt{UNKNOWN} after convergence (Section~\ref{sec:method:ifc}). Table~\ref{tab:cond-eval} gives   %
the resulting rules for AND and OR, which follow Kleene's strong
three-valued logic~\cite{kleen}. A \texttt{NO} on either side
decides AND, and a \texttt{YES} on either side decides OR. Without a
deciding value, the result is \texttt{UNKNOWN} if either operand is
\texttt{UNKNOWN}.

Both connectives are monotone in the information order, in which
\texttt{UNKNOWN} lies below \texttt{YES} and \texttt{NO}. Settling an
operand can turn an \texttt{UNKNOWN} result into \texttt{YES} or
\texttt{NO}, but never change one settled value into the other. This is
the property used in Section~\ref{sec:method:ifc} that no value       %
resolved later contradicts one resolved earlier.

\subsection{Path Verification Agent}
\label{appendix:p4}

This stage adjudicates each candidate path emitted by the traversal
stage as a true or false positive via one separate LLM call.
Verdicts are produced without cross-path context and are therefore
mutually independent and order-insensitive. Unlike prompts P1--P3, this prompt is assembled per call from a shared
evidence header, a hint block selected by the path's category, and a
fixed output specification. Only the hint block differs between path
categories, and several categories share one hint block
(Table~\ref{tab:pattern-families}).                                          %

\begin{table}[h]
\centering
\footnotesize
\caption{Pattern families (\texttt{path\_type}) and the verifier hint
group each uses. Each \texttt{path\_type} corresponds to one pattern family of the
graph traversal (Section~\ref{sec:method:acpg}).}
\label{tab:pattern-families}
\begin{tabular}{llp{0.42\linewidth}}
\toprule
\textbf{Group} & \textbf{\texttt{path\_type}} & \textbf{Vulnerability class} \\
\midrule
1   & \texttt{taint\_chain}      & Secret / Credential Exposure \\
2a  & \texttt{exec\_chain}       & Code Execution / RCE \\
2b  & \texttt{ipi\_chain}        & Indirect Prompt Injection \\
2c  & \texttt{subprocess\_chain} & Secret / Credential Exposure \\
3a  & \texttt{idor\_chain}       & Access Control / IDOR \\
3b  & \texttt{confused\_deputy}  & LLM confused deputy / escalation \\
3c  & \texttt{secret\_return}    & Secret / Credential Exposure \\
3d  & \texttt{store\_poison}     & Persistent store poisoning \\
4   & \texttt{ext\_chain}        & Other \\
\bottomrule
\end{tabular}
\end{table}

Each hint block states explicit criteria for classifying a path as a
true or false positive within its class. Each block is written once per vulnerability class, not per agent, and refers only to ACPG properties and sink kinds. Group~4 covers paths built from extended properties, which have no fixed vulnerability class. Its hints are therefore illustrative, and the verifier may judge from the source code. The verifier also assigns each path its family in Table~\ref{tab:findings-summary}. These families describe the vulnerability, so they do not map one-to-one onto the pattern families in Table~\ref{tab:pattern-families}. Output is a fixed six-field JSON object.

\begin{quote}\footnotesize\ttfamily
"You are a security analyst reviewing candidate vulnerability paths detected
by a static analysis pipeline on LLM-based agent code. Your task: decide if
each path represents a real vulnerability (true positive) or a false
positive. Base ALL decisions strictly on the evidence provided. Do not
invent anything. CRITICAL: You must respond with ONLY a JSON object. No
prose, no markdown, no explanation. Your entire response must be a single
valid JSON object starting with \{ and ending with \}."
\normalfont\footnotesize(Prompt P4: path verification system prompt)
\end{quote}

The per-call user message opens with a shared evidence header
combining component-level static analysis evidence, the resolved
property map, and, for certain path categories, additional contextual
metadata, followed by the verbatim source of the implicated files.

\begin{lstlisting}[style=promptcode, caption={Prompt P4: shared evidence header (per-path).}, label={lst:p4_header}]
## Path under review ({path_index} of {n_total})
path_type   : {path_type}
path        : {path_str}
source_node : {source_node}
sink_node   : {sink_node}
tier        : {tier}
{extra_meta}
## Per-component static analysis evidence
{component_evidence_str}

## Source files
{source_files_str}
\end{lstlisting}

A representative subset of Group~1's hint block is mentioned below,
illustrating how each entry distinguishes true from false positives
using evidence specific to the sink involved.

\begin{lstlisting}[style=promptcode, caption={Prompt P4: Group~1 hint block excerpt: credential exfiltration.}, label={lst:p4_phase1}]
## Group 1 - Credential exfil path
Secret (credential/env var) flowing to an external boundary.

sink_node = public_output:*  (LLM-mediated exfil)
  TP: secret_in_tool_return=YES, highest_taint_return=SECRET before llm_node
  FP: only main->sink, no intermediate tool reads secret independently

sink_node = http_post:* or http_request:*  (out-of-band HTTP exfil)
  TP: external_communication=YES + may_leak_secret=YES + reads_secret=YES
  Note: secret_in_tool_return=NO is expected here (secret leaves via HTTP, not via the LLM); do not treat it as an FP signal

sink_node = http_post:* with file: conduit in path  (file-bridge exfil)
  One component writes secret to file; another reads and POSTs it

  [...]

\end{lstlisting}

Every call terminates with the same output specification, including the verdict and a severity rating. The verifier runs on Sonnet 4.6 at temperature 0.

\subsection{LLM-Only Baseline Strategies}
\label{appendix:llm-only-prompts}

This section shows the core prompt of each LLM-only strategy in
Section~\ref{sec:eval:setup}. Every strategy receives the source code with
either the definitions of the 19 core properties (RQ1,
Table~\ref{tab:core_properties}) or the 7-fact schema for threat chain
reconstruction (Section~\ref{sec:eval:setup}). Ground truth is never
shown to the model. Agentic and Five-Round run in both RQs and differ only
in the output format they request. The other five strategies are each
specific to one RQ.

\textbf{Agentic (RQ1, RQ2).} This strategy invokes \texttt{claude -p} with
the \texttt{Read} tool, so the model explores the source folder on its own
and reads files on demand. The system prompt fixes a security-analyst
persona and defines the task's key terms, and the task prompt asks for
exactly one JSON object with no prose. That object is the 19-property
template for RQ1 and an array of 7-fact chains for RQ2. Only the closing
output-format block of the prompt changes between the two. Agentic runs at
the provider's default settings.
\begin{quote}\footnotesize\ttfamily
"You are a senior security engineer conducting a focused [...] analysis
of an AI agent codebase. You have access to the Read tool to explore
files on demand. Only read files within the current directory. [...]
Investigate this AI agent codebase for
vulnerabilities. Use the Read tool to explore source files and trace the
attack chain." \normalfont\footnotesize(RQ2 system prompt)
\end{quote}

\textbf{Five-Round / Chain Tracer (RQ1 / RQ2).} This strategy simulates a
static analyzer over five sequential rounds under one taint-analyst system
prompt, and each round receives the source and every earlier round's
output. The rounds are (R1) a function inventory of reads, writes, and
calls, (R2) taint source identification, (R3) sink identification, (R4)
source-to-sink chain tracing, and (R5) a structured verdict. R1 to R4 are
shared verbatim between the two tasks. R5 asks for the property template
in RQ1 and, in RQ2, for a JSON array of chains, each listing the seven
facts with a confidence level and notes.
\begin{quote}\footnotesize\ttfamily
"ROUND 1 -- FUNCTION INVENTORY. For every function defined in these
files, produce a concise entry [...]. Be exhaustive. Include private
helpers (underscore-prefixed). Do not skip any function. Do not make
security judgements yet -- just map the code." \normalfont\footnotesize(R1 prompt, verbatim across RQ1/RQ2)
\end{quote}

\textbf{One-shot (RQ1).} This strategy makes a single call with the full
source and the property definitions in context, with no iterative
structure. It isolates how much the five-round scaffold adds.
\begin{quote}\footnotesize\ttfamily
"You are a static security analysis tool. Your task is to fill a
structured capability template for each component (function/tool)
in an AI agent codebase. You must reason purely from the source code
provided." \normalfont\footnotesize(system prompt)
\end{quote}

\textbf{Persona (RQ1).} This strategy replaces the analysis rounds with five analyst personas. Four of them analyze the same source independently: Data Flow, Supply Chain Auditor, IFC, and Red Team. Each one looks at the code from its own perspective. The fifth, the Synthesizer, merges their four analyses into the property template. It marks a property \texttt{YES} if any persona flagged it.
\begin{quote}\footnotesize\ttfamily
"You are a red team security researcher. [...] What is the most subtle
way data could be exfiltrated? Consider: values embedded in data
structures, HTTP calls hidden in helper chains, side-channel writes to
files. Describe every potential exfiltration path you find, even if it
requires a compromised component." \normalfont\footnotesize(Red Team persona, round 4)
\end{quote}
\begin{quote}\footnotesize\ttfamily
"You have received 4 independent analyses of this codebase from
different security perspectives. Synthesize all findings into the final
property template. If ANY analyst identified a taint path or secret
access for a component, [...], mark the relevant property YES."
\normalfont\footnotesize(Synthesizer, round 5)
\end{quote}

\textbf{Verifier (RQ1).} Instance A runs Five-Round. A fresh Instance B
sees only the source and Instance A's JSON, and must find code evidence
against every \texttt{NO}. The challenge is one-directional, because a
missed capability is the costlier error, while a spurious \texttt{YES} is
cheap to filter. The final value is \texttt{YES} if either instance says
\texttt{YES}.
\begin{quote}\footnotesize\ttfamily
"You are a second, independent static security analyst performing
adversarial review. [...] Your sole task is to CHALLENGE every property
marked "NO" in Instance A's output. [...] Search the source code for ANY
code evidence that contradicts it." \normalfont\footnotesize(Instance B system prompt)
\end{quote}

\textbf{Adversarial (RQ1).} This strategy is the single-instance analogue
of Verifier. After a Five-Round run, a sixth round asks the same instance to back every \texttt{NO} with a line of code or flip it to \texttt{YES}.
\begin{quote}\footnotesize\ttfamily
"Your task is to challenge every property you marked NO. [...] Try to
find ANY line of code that proves taint DOES reach that property. [...]
If you cannot find a concrete line of code that proves the taint is
absent [...], change the property to YES. Be adversarial: assume taint
travels further than you initially thought." \normalfont\footnotesize(Round 6 prompt)
\end{quote}

\textbf{Code Reviewer (RQ2).} This strategy runs in two steps, both with
the full source in context. An open-ended security review comes first,
and a second step extracts its findings into the 7-fact schema.
\begin{quote}\footnotesize\ttfamily
"You are a senior security engineer conducting a focused threat chain
analysis of an AI agent codebase." \normalfont\footnotesize(Step~1 system prompt)
\end{quote}

\subsection{RQ1 Additional Results}\label{app:rq1_additional}
Table~\ref{tab:rq1_f1_comparison} reports per-agent macro-F1 scores for
ACaM-SA (static analysis only), LLM-only (best prompting strategy and model), and
the full \sysname{} pipeline. The aggregates in Table~\ref{tab:rq1_aggregate} (Section~\ref{sec:eval:rq1}) first average each group, base and vulnerable, and then average the two group means.

\begin{table}[t]
\centering
\caption{Per-agent macro-F1 across ACaM-SA (static analysis only),
LLM-only (best prompting strategy and model), and ACaM (full pipeline).}
\label{tab:rq1_f1_comparison}
\scriptsize
\setlength{\tabcolsep}{4pt}
\begin{tabular}{@{}lccc@{}}
\toprule
\textbf{Agent} & \textbf{ACaM-SA} & \textbf{LLM-only} & \textbf{ACaM} \\
\midrule
\multicolumn{4}{@{}l}{\textit{Base agents}} \\
B1 & 0.972 & 0.893 & \textbf{1.000} \\
B2 & 0.966 & 0.666 & \textbf{0.975} \\
B3 & 0.925 & 0.756 & \textbf{0.937} \\
B4 & 0.996 & 0.900 & \textbf{1.000} \\
B5 & 0.997 & 0.854 & \textbf{1.000} \\
B6 & 0.997 & 0.817 & \textbf{1.000} \\
B7 & 0.997 & 0.786 & \textbf{1.000} \\
B8 & 0.956 & 0.850 & \textbf{0.962} \\
B9 & 0.664 & 0.720 & \textbf{0.887} \\
B10 & 0.968 & 0.856 & \textbf{0.969} \\
\midrule
\multicolumn{4}{@{}l}{\textit{Vulnerable agents}} \\
B1\_A1\_A2 & 0.983 & 0.777 & \textbf{1.000} \\
B2\_A1 & 0.989 & 0.610 & \textbf{0.995} \\
B3\_A3\_A2 & 0.949 & 0.762 & \textbf{0.993} \\
B4\_A1 & 0.692 & 0.917 & \textbf{1.000} \\
B5\_A1 & 0.984 & 0.913 & \textbf{0.987} \\
B5\_A2 & 0.986 & 0.945 & \textbf{1.000} \\
B6\_A1 & 0.987 & 0.827 & \textbf{0.989} \\
B6\_A2 & 0.997 & 0.843 & \textbf{1.000} \\
B7\_A1 & 0.987 & 0.883 & \textbf{0.989} \\
B7\_A3\_A2 & 0.985 & 0.893 & \textbf{0.991} \\
B8\_A1 & 0.970 & 0.805 & \textbf{0.983} \\
B9\_A1 & 0.428 & 0.586 & \textbf{0.983} \\
B9\_A2 & 0.985 & 0.817 & \textbf{1.000} \\
B10\_A1 & 0.994 & 0.872 & \textbf{0.996} \\
\bottomrule
\end{tabular}
\par\vspace{3pt}
\parbox{\linewidth}{\footnotesize\raggedright
Base agents are anonymized as B1 to B10 pending coordinated disclosure; we will restore their names once fixes are released.
Variant suffixes: A1 = secret returned to the LLM (ASI02, ASI06);
A2 = hidden out-of-band channel, such as an HTTP POST (ASI02, ASI07); A3 = indirection through shared
state or a subprocess. Combined suffixes inject both classes.}
\end{table}
\subsection{RQ2 Additional Results}\label{app_rq2_additional}
This subsection provides detailed results for the RQ2 Part~1 (Section~\ref{sec:eval:rq2:ds1}) threat chain
reconstruction phase. Table~\ref{tab:rq2_chain_scores} reports the per-agent
score breakdown.

\begin{table}[!htb]
\centering
\caption{Threat chain reconstruction scores across the 14 injected vulnerable
variants.
LLM columns report the best prompting strategy per model; Pysa runs
without (NS) and with (WS) taint-in-taint-out stubs. }
\label{tab:rq2_chain_scores}
\scriptsize
\setlength{\tabcolsep}{2pt}
\begin{tabular}{@{}lcccccc@{}}
\toprule
 & & \multicolumn{2}{c}{\textbf{Pysa}} & & \multicolumn{2}{c}{\textbf{LLM}} \\
\cmidrule(lr){3-4}\cmidrule(lr){6-7}
\textbf{Agent} & \textbf{ACaM} & \textbf{NS} & \textbf{WS} & \textbf{Semgrep} & \textbf{Son.} & \textbf{Opus} \\
\midrule
B1\_A1\_A2 & \textbf{1.000} & 0.000 & 0.786 & 0.000 & 0.929 & 0.929 \\
B2\_A1 & \textbf{1.000} & 0.286 & 0.357 & 0.143 & 0.929 & \textbf{1.000} \\
B3\_A3\_A2 & \textbf{1.000} & 0.000 & 0.000 & 0.714 & 0.929 & 0.929 \\
B4\_A1 & \textbf{0.929} & 0.357 & 0.357 & 0.000 & 0.857 & 0.714 \\
B5\_A1 & \textbf{1.000} & 0.286 & 0.357 & 0.071 & 0.857 & 0.857 \\
B5\_A2 & \textbf{1.000} & 0.929 & 0.143 & 0.857 & \textbf{1.000} & 0.857 \\
B6\_A1 & \textbf{1.000} & 0.286 & 0.357 & 0.071 & 0.857 & 0.857 \\
B6\_A2 & \textbf{1.000} & 0.929 & 0.143 & 0.786 & 0.857 & \textbf{1.000} \\
B7\_A1 & \textbf{1.000} & 0.286 & 0.000 & 0.071 & 0.857 & 0.857 \\
B7\_A3\_A2 & \textbf{1.000} & 0.000 & 0.143 & 0.000 & 0.857 & \textbf{1.000} \\
B8\_A1 & \textbf{1.000} & 0.000 & 0.000 & 0.071 & 0.929 & 0.929 \\
B9\_A1 & \textbf{0.857} & 0.000 & 0.000 & 0.071 & \textbf{0.857} & 0.714 \\
B9\_A2 & \textbf{1.000} & 0.929 & 0.000 & 0.857 & \textbf{1.000} & 0.857 \\
B10\_A1 & \textbf{0.929} & 0.500 & 0.500 & 0.000 & \textbf{0.929} & \textbf{0.929} \\
\midrule
\textit{Mean} & \textbf{0.980} & 0.342 & 0.225 & 0.265 & 0.903 & 0.888 \\
\bottomrule
\end{tabular}
\par\vspace{3pt}
\parbox{\linewidth}{\footnotesize\raggedright
Agent IDs follow Table~\ref{tab:rq1_f1_comparison}.
Variant suffixes: A1 = secret returned to the LLM (ASI02, ASI06);
A2 = hidden out-of-band channel, such as an HTTP POST (ASI02, ASI07); A3 = indirection through shared
state or a subprocess. Combined suffixes inject both classes, as one chain or two.}
\end{table}

%% file: main.bib
@article{code_property_graph,
  title={Modeling and Discovering Vulnerabilities with Code Property Graphs},
  author={Fabian Yamaguchi and Nico Golde and Daniel Arp and Konrad Rieck},
  journal={2014 IEEE Symposium on Security and Privacy},
  year={2014},
  pages={590-604},
  url={https://api.semanticscholar.org/CorpusID:2231082}
}

@article{kahn_topo,
author = {Kahn, A. B.},
title = {Topological sorting of large networks},
year = {1962},
issue_date = {Nov. 1962},
publisher = {Association for Computing Machinery},
address = {New York, NY, USA},
volume = {5},
number = {11},
issn = {0001-0782},
url = {https://doi.org/10.1145/368996.369025},
doi = {10.1145/368996.369025},
journal = {Commun. ACM},
month = nov,
pages = {558–562},
numpages = {5}
}

@misc{scalpel,
      title={Scalpel: The Python Static Analysis Framework}, 
      author={Li Li and Jiawei Wang and Haowei Quan},
      year={2022},
      eprint={2202.11840},
      archivePrefix={arXiv},
      primaryClass={cs.SE},
      url={https://arxiv.org/abs/2202.11840}, 
}

@misc{stub-bugstone,
      title={One Bug, Hundreds Behind: LLMs for Large-Scale Bug Discovery}, 
      author={Qiushi Wu and Yue Xiao and Dhilung Kirat and Kevin Eykholt and Jiyong Jang and Douglas Lee Schales},
      year={2025},
      eprint={2510.14036},
      archivePrefix={arXiv},
      primaryClass={cs.SE},
      url={https://arxiv.org/abs/2510.14036}, 
}

@misc{stub-llmdepgraph,
      title={AgentArmor: Enforcing Program Analysis on Agent Runtime Trace to Defend Against Prompt Injection}, 
      author={Peiran Wang and Yang Liu and Yunfei Lu and Yifeng Cai and Hongbo Chen and Qingyou Yang and Jie Zhang and Jue Hong and Ye Wu},
      year={2025},
      eprint={2508.01249},
      archivePrefix={arXiv},
      primaryClass={cs.CR},
      url={https://arxiv.org/abs/2508.01249}, 
}

@misc{forge_jha,
      title={Formal Policy Enforcement for Real-World Agentic Systems}, 
      author={Nils Palumbo and Sarthak Choudhary and Jihye Choi and Guy Amir and Prasad Chalasani and Somesh Jha},
      year={2026},
      eprint={2602.16708},
      archivePrefix={arXiv},
      primaryClass={cs.CR},
      url={https://arxiv.org/abs/2602.16708}, 
}

@misc{stub-pysa,
  title        = {{Pysa}: {Python} Static Analyzer},
  author       = {{Meta Platforms, Inc.}},
  howpublished = {\url{https://pyre-check.org/docs/pysa-basics/}},
  note         = {Accessed: 2026-08-13},
  year         = {2026}
}

@misc{stub-codeql,
  title        = {{CodeQL}: {Python} Query Support},
  author       = {{GitHub, Inc.}},
  howpublished = {\url{https://github.com/github/codeql/tree/main/python}},
  note         = {Accessed: 2026-08-13},
  year         = {2026}
}

@misc{stub-semgrep,
  title        = {{Semgrep}: Lightweight Static Analysis for Many Languages},
  author       = {{Semgrep, Inc.}},
  howpublished = {\url{https://github.com/semgrep/semgrep}},
  note         = {Accessed: 2026-08-13},
  year         = {2026}
}

@misc{stub-stridegpt,
  title        = {{STRIDE GPT}: An AI-Powered Threat Modeling Tool},
  author       = {Adams, Matt},
  howpublished = {\url{https://github.com/mrwadams/stride-gpt}},
  note         = {Accessed: 2026-08-13},
  year         = {2026}
}

@misc{stub-iris,
      title={IRIS: LLM-Assisted Static Analysis for Detecting Security Vulnerabilities}, 
      author={Ziyang Li and Saikat Dutta and Mayur Naik},
      year={2025},
      eprint={2405.17238},
      archivePrefix={arXiv},
      primaryClass={cs.CR},
      url={https://arxiv.org/abs/2405.17238}, 
}

@inproceedings{stub-ifc,
author = {Ntousakis, Grigoris and Stephen, Julian James and Le, Michael V. and Chukkapalli, Sai Sree Laya and Taylor, Teryl and Molloy, Ian M. and Araujo, Frederico},
title = {Securing MCP-based Agent Workflows},
year = {2025},
isbn = {9798400722059},
publisher = {Association for Computing Machinery},
address = {New York, NY, USA},
url = {https://doi.org/10.1145/3766882.3767177},
doi = {10.1145/3766882.3767177},
booktitle = {Proceedings of the 4th Workshop on Practical Adoption Challenges of ML for Systems},
pages = {50–55},
numpages = {6},
location = {Seoul, Republic of Korea},
series = {PACMI '25}
}

@misc{sok_jailbreak,
      title={SoK: Rethinking Jailbreaking in the Era of Agentic AI: Attacks, Defenses, and Practical Consideration}, 
      author={Md Jueal Mia and Yanzhao Wu and Selcuk Uluagac and M. Hadi Amini},
      year={2026},
      eprint={2609.12413},
      archivePrefix={arXiv},
      primaryClass={cs.AI},
      url={https://arxiv.org/abs/2609.12413}, 
}

@misc{stub-camel,
      title={Defeating Prompt Injections by Design}, 
      author={Edoardo Debenedetti and Ilia Shumailov and Tianqi Fan and Jamie Hayes and Nicholas Carlini and Daniel Fabian and Christoph Kern and Chongyang Shi and Andreas Terzis and Florian Tramèr},
      year={2025},
      eprint={2503.18813},
      archivePrefix={arXiv},
      primaryClass={cs.CR},
      url={https://arxiv.org/abs/2503.18813}, 
}

@misc{stub-christodorescu,
      title={Agent Security is a Systems Problem}, 
      author={Mihai Christodorescu and Earlence Fernandes and Ashish Hooda and Somesh Jha and Johann Rehberger and Kamalika Chaudhuri and Xiaohan Fu and Khawaja Shams and Guy Amir and Jihye Choi and Sarthak Choudhary and Nils Palumbo and Andrey Labunets and Nishit V. Pandya},
      year={2026},
      eprint={2605.18991},
      archivePrefix={arXiv},
      primaryClass={cs.CR},
      url={https://arxiv.org/abs/2605.18991}, 
}

@misc{ipi,
      title={Not what you've signed up for: Compromising Real-World LLM-Integrated Applications with Indirect Prompt Injection}, 
      author={Kai Greshake and Sahar Abdelnabi and Shailesh Mishra and Christoph Endres and Thorsten Holz and Mario Fritz},
      year={2023},
      eprint={2302.12173},
      archivePrefix={arXiv},
      primaryClass={cs.CR},
      url={https://arxiv.org/abs/2302.12173}, 
}

@misc{nvd2025cve68664,
  title        = {{CVE-2025-68664}},
  howpublished = {National Vulnerability Database},
  year         = {2025},
  url          = {https://nvd.nist.gov/vuln/detail/CVE-2025-68664},
  note         = {Accessed: 2026-09-03}
}

@misc{nvd2024cve21513,
  title        = {{CVE-2024-21513}},
  howpublished = {National Vulnerability Database},
  year         = {2024},
  url          = {https://nvd.nist.gov/vuln/detail/CVE-2024-21513},
  note         = {Accessed: 2026-09-03}
}

@inproceedings{taint_analysis_1,
author = {Schwartz, Edward J. and Avgerinos, Thanassis and Brumley, David},
title = {All You Ever Wanted to Know about Dynamic Taint Analysis and Forward Symbolic Execution (but Might Have Been Afraid to Ask)},
year = {2010},
isbn = {9780769540351},
publisher = {IEEE Computer Society},
address = {USA},
url = {https://doi.org/10.1109/SP.2010.26},
doi = {10.1109/SP.2010.26},
booktitle = {Proceedings of the 2010 IEEE Symposium on Security and Privacy},
pages = {317–331},
numpages = {15},
series = {SP '10}
}

@inproceedings{taint_analysis_2,
  title={Pixy : A Static Analysis Tool for Detecting Web Application Vulnerabilities ( Technical Report )},
  author={Nenad Jovanovi{\'c} and Christopher Kr{\"u}gel and Engin Kirda},
  year={2006},
  url={https://api.semanticscholar.org/CorpusID:262089347}
}

@article{taint_analysis_3,
author = {Arzt, Steven and Rasthofer, Siegfried and Fritz, Christian and Bodden, Eric and Bartel, Alexandre and Klein, Jacques and Le Traon, Yves and Octeau, Damien and McDaniel, Patrick},
title = {FlowDroid: precise context, flow, field, object-sensitive and lifecycle-aware taint analysis for Android apps},
year = {2014},
issue_date = {June 2014},
publisher = {Association for Computing Machinery},
address = {New York, NY, USA},
volume = {49},
number = {6},
issn = {0362-1340},
url = {https://doi.org/10.1145/2666356.2594299},
doi = {10.1145/2666356.2594299},
journal = {SIGPLAN Not.},
month = jun,
pages = {259–269},
numpages = {11}
}

@misc{threatModel,
      title={ThreatModeling-LLM: Automating Threat Modeling using Large Language Models for Banking System}, 
      author={Tingmin Wu and Shuiqiao Yang and Shigang Liu and David Nguyen and Seung Jang and Alsharif Abuadbba},
      year={2025},
      eprint={2411.17058},
      archivePrefix={arXiv},
      primaryClass={cs.CR},
      url={https://arxiv.org/abs/2411.17058}, 
}

@inproceedings{neurosym1,
  title={LineBreaker: Finding Token-Inconsistency Bugs with Large Language Models},
  author={Chen, Hongbo and Zhang, Yifan and Han, Xing and Mao, Tianhao and Rong, Huanyao and Zhang, Yuheng and Wang, XiaoFeng and Xing, Luyi and Chen, Xun and Zhang, Hang},
  booktitle={2025 40th IEEE/ACM International Conference on Automated Software Engineering (ASE)},
  pages={893--905},
  year={2025},
  organization={IEEE}
}

@inproceedings{neurosym2,
  title={Scope: Evaluating llms for software vulnerability detection},
  author={Gon{\c{c}}alves, Jos{\'e} and Dias, Tiago and Maia, Eva and Pra{\c{c}}a, Isabel},
  booktitle={International Symposium on Distributed Computing and Artificial Intelligence},
  pages={34--43},
  year={2024},
  organization={Springer}
}

@inproceedings{neurosym3,
  title={Assessing the effectiveness of llms in android application vulnerability analysis},
  author={Kouliaridis, Vasileios and Karopoulos, Georgios and Kambourakis, Georgios},
  booktitle={International Conference on Attacks and Defenses for Internet-of-Things},
  pages={139--154},
  year={2024},
  organization={Springer}
}

@inproceedings{neurosym4,
  title={From Large to Mammoth: A Comparative Evaluation of Large Language Models in Vulnerability Detection.},
  author={Lin, Jie and Mohaisen, David and others},
  booktitle={NDSS},
  year={2025}
}

@inproceedings{neurosym5,
  title={Exploring $\{$ChatGPT's$\}$ capabilities on vulnerability management},
  author={Liu, Peiyu and Liu, Junming and Fu, Lirong and Lu, Kangjie and Xia, Yifan and Zhang, Xuhong and Chen, Wenzhi and Weng, Haiqin and Ji, Shouling and Wang, Wenhai},
  booktitle={33rd USENIX Security Symposium (USENIX Security 24)},
  pages={811--828},
  year={2024}
}

@inproceedings{neurosym6,
  title={Llms cannot reliably identify and reason about security vulnerabilities (yet?): A comprehensive evaluation, framework, and benchmarks},
  author={Ullah, Saad and Han, Mingji and Pujar, Saurabh and Pearce, Hammond and Coskun, Ayse and Stringhini, Gianluca},
  booktitle={2024 IEEE symposium on security and privacy (SP)},
  pages={862--880},
  year={2024},
  organization={IEEE}
}

@article{neurosym7,
  title={Multitask-based evaluation of open-source llm on software vulnerability},
  author={Yin, Xin and Ni, Chao and Wang, Shaohua},
  journal={IEEE Transactions on Software Engineering},
  volume={50},
  number={11},
  pages={3071--3087},
  year={2024},
  publisher={IEEE}
}

@misc{fides,
      title={Securing AI Agents with Information-Flow Control}, 
      author={Manuel Costa and Boris Köpf and Aashish Kolluri and Andrew Paverd and Mark Russinovich and Ahmed Salem and Shruti Tople and Lukas Wutschitz and Santiago Zanella-Béguelin},
      year={2025},
      eprint={2505.23643},
      archivePrefix={arXiv},
      primaryClass={cs.CR},
      url={https://arxiv.org/abs/2505.23643}, 
}

@misc{isolategpt,
      title={IsolateGPT: An Execution Isolation Architecture for LLM-Based Agentic Systems}, 
      author={Yuhao Wu and Franziska Roesner and Tadayoshi Kohno and Ning Zhang and Umar Iqbal},
      year={2025},
      eprint={2403.04960},
      archivePrefix={arXiv},
      primaryClass={cs.CR},
      url={https://arxiv.org/abs/2403.04960}, 
}

@misc{nvd2025cve55284,
  title        = {{CVE-2025-55284}},
  howpublished = {National Vulnerability Database},
  year         = {2025},
  url          = {https://nvd.nist.gov/vuln/detail/CVE-2025-55284},
  note         = {Accessed: 2026-09-03}
}

@misc{owasp2026agentic,
  author       = {{OWASP Foundation}},
  title        = {OWASP Top 10 For Agentic
Applications 2026},
  year         = {2026},
  howpublished = {\url{https://genai.owasp.org/resource/owasp-top-10-for-agentic-applications-for-2026/}},
  note         = {Accessed: 2026-09-05}
}

@misc{stub-dcisurvey,
      title={Description-Code Inconsistency in Real-world MCP Servers: Measurement, Detection, and Security Implications},
      author={Yutao Shi and Xiaohan Zhang and Xiangjing Zhang and Xihua Shen and Hui Ouyang and Huming Qiu and Mi Zhang and Min Yang},
      year={2026},
      eprint={2606.04769},
      archivePrefix={arXiv},
      primaryClass={cs.CR},
      url={https://arxiv.org/abs/2606.04769},
}

@misc{dci_2,
      title={Don't believe everything you read: Understanding and Measuring MCP Behavior under Misleading Tool Descriptions}, 
      author={Zhihao Li and Boyang Ma and Xuelong Dai and Minghui Xu and Yue Zhang and Biwei Yan and Kun Li},
      year={2026},
      eprint={2602.03580},
      archivePrefix={arXiv},
      primaryClass={cs.CR},
      url={https://arxiv.org/abs/2602.03580}, 
}

@article{progent,
  title={Progent: Securing AI agents with privilege control},
  author={Shi, Tianneng and He, Jingxuan and Wang, Zhun and Li, Hongwei and Wu, Linyu and Guo, Wenbo and Song, Dawn},
  journal={arXiv preprint arXiv:2504.11703},
  year={2025}
}

@article{agent_sec_1,
  title={ARGUS: Defending LLM agents against context-aware prompt injection},
  author={Weng, Shihao and Feng, Yang and Zhang, Jinrui and Xie, Xiaofei and Yu, Jiongchi and Liu, Jia},
  journal={arXiv preprint arXiv:2605.03378},
  year={2026}
}

@article{agent_sec_2,
  title={Efficient and Sound Probabilistic Verification for AI Agents},
  author={Solko-Breslin, Alaia and Mudrakarta, Pramod Kaushik and Christodorescu, Mihai and Jha, Somesh and Dvijotham, Krishnamurthy Dj},
  journal={arXiv preprint arXiv:2606.20510},
  year={2026}
}

@misc{agent_sec_3,
      title={Systems Security Foundations for Agentic Computing}, 
      author={Mihai Christodorescu and Earlence Fernandes and Ashish Hooda and Somesh Jha and Johann Rehberger and Kamalika Chaudhuri and Xiaohan Fu and Khawaja Shams and Guy Amir and Jihye Choi and Sarthak Choudhary and Nils Palumbo and Andrey Labunets and Nishit V. Pandya},
      year={2026},
      eprint={2512.01295},
      archivePrefix={arXiv},
      primaryClass={cs.CR},
      url={https://arxiv.org/abs/2512.01295}, 
}

@article{agent_Sec_5,
  title={System-level defense against indirect prompt injection attacks: An information flow control perspective},
  author={Wu, Fangzhou and Cecchetti, Ethan and Xiao, Chaowei},
  journal={arXiv preprint arXiv:2409.19091},
  year={2024}
}

@inproceedings {adaptive_attack,
author = {Milad Nasr and Nicholas Carlini and Chawin Sitawarin and Sander V. Schulhoff and Jamie Hayes and Michael Ilie and Juliette Pluto and Shuang Song and Harsh Chaudhari and Ilia Shumailov and Abhradeep Guha Thakurta and Kai Yuanqing Xiao and Andreas Terzis and Florian Tram{\`e}r},
title = {The Attacker Moves Second: Stronger Adaptive Attacks Bypass Defenses Against {LLM} Jailbreaks and Prompt Injections},
booktitle = {35th USENIX Security Symposium (USENIX Security 26)},
year = {2026},
isbn = {978-1-939133-58-8},
address = {Baltimore, MD},
pages = {1467--1486},
url = {https://www.usenix.org/conference/usenixsecurity26/presentation/nasr},
publisher = {USENIX Association},
month = aug
}

@misc{capa_launder_file_system,
      title={My Credentials, Please: Detecting Filesystem Capability Laundering in MCP Servers at Scale}, 
      author={Zeqian Huang and Mingyue Tang and Aonan Guan and Qiming Sun and Pengzhou He},
      year={2026},
      eprint={2610.22301},
      archivePrefix={arXiv},
      primaryClass={cs.CR},
      url={https://arxiv.org/abs/2610.22301}, 
}

@misc{agent_sec_7,
      title={DRIFT: Dynamic Rule-Based Defense with Injection Isolation for Securing LLM Agents}, 
      author={Hao Li and Xiaogeng Liu and Hung-Chun Chiu and Dianqi Li and Ning Zhang and Chaowei Xiao},
      year={2026},
      eprint={2506.12104},
      archivePrefix={arXiv},
      primaryClass={cs.CR},
      url={https://arxiv.org/abs/2506.12104}, 
}

@article{agent_sec_8,
  title={Ace: A security architecture for llm-integrated app systems},
  author={Li, Evan and Mallick, Tushin and Rose, Evan and Robertson, William and Oprea, Alina and Nita-Rotaru, Cristina},
  journal={arXiv preprint arXiv:2504.20984},
  year={2025}
}

@article{agent_sec_9,
  title={Aligning Provenance with Authorization: A Dual-Graph Defense for LLM Agents},
  author={Wang, Peiran and Li, Ying and Tian, Yuan},
  journal={arXiv preprint arXiv:2605.26497},
  year={2026}
}

@inproceedings{agent_sec_10,
  title={Secalign: Defending against prompt injection with preference optimization},
  author={Chen, Sizhe and Zharmagambetov, Arman and Mahloujifar, Saeed and Chaudhuri, Kamalika and Wagner, David and Guo, Chuan},
  booktitle={Proceedings of the 2025 ACM SIGSAC Conference on Computer and Communications Security},
  pages={2833--2847},
  year={2025}
}

@misc{agent_sec_11,
      title={ceLLMate: Sandboxing Browser AI Agents}, 
      author={Luoxi Meng and Henry Feng and Ilia Shumailov and Earlence Fernandes},
      year={2026},
      eprint={2512.12594},
      archivePrefix={arXiv},
      primaryClass={cs.CR},
      url={https://arxiv.org/abs/2512.12594}, 
}

@inproceedings{kleen,
author = {Sagiv, Mooly and Reps, Thomas and Wilhelm, Reinhard},
title = {Parametric shape analysis via 3-valued logic},
year = {1999},
isbn = {1581130953},
publisher = {Association for Computing Machinery},
address = {New York, NY, USA},
url = {https://doi.org/10.1145/292540.292552},
doi = {10.1145/292540.292552},
booktitle = {Proceedings of the 26th ACM SIGPLAN-SIGACT Symposium on Principles of Programming Languages},
pages = {105–118},
numpages = {14},
location = {San Antonio, Texas, USA},
series = {POPL '99}
}
